\documentclass[12pt]{JHEP3}
\usepackage{amsmath,amssymb,epsfig,bm,amsfonts,yfonts}

\newcommand{\be}{\begin{equation}}
\newcommand{\ee}{\end{equation}}
\newcommand{\beqa}{\begin{eqnarray}}
\newcommand{\eeqa}{\end{eqnarray}}
 \newcommand{\bn}{\begin{enumerate}}
\newcommand{\en}{\end{enumerate}}

\def\bra#1{\left\langle #1\right|}
\def\eeq{\end{equation}}

\def\ket#1{\left| #1\right\rangle}

\def\Tr{\mathop{\rm Tr}}

\relax

\title{Bulk Viscosity and Cavitation in Boost-Invariant Hydrodynamic Expansion}

\author{Krishna Rajagopal${}^{\,1}$ and Nilesh Tripuraneni${}^{\,2}$\\
\vspace{0.1in}

${}^{\,1}$Center for Theoretical Physics, 
Massachusetts Institute
of Technology,
Cambridge, MA 02139, USA
\vspace{0.1in}

${}^{\,2}$Clovis West High School,
Fresno, CA 93720, USA
\vspace{0.1in}

E-mail addresses: {\tt krishna@mit.edu, nileshtrip@gmail.com}
}

\abstract{We solve second order relativistic hydrodynamics equations for a boost-invariant $1+1$-dimensional expanding fluid with an equation of state taken from lattice calculations of the thermodynamics of strongly coupled quark-gluon plasma.  We investigate the dependence of the energy density as a function of proper time on the values of the shear viscosity $\eta$, the bulk viscosity $\zeta$, and second order coefficients, confirming that large changes in the values of the latter have negligible effects.   Varying the shear viscosity between zero and a few times $s/4\pi$, with $s$ the entropy density, has significant effects, as expected based on other studies.   Introducing a nonzero bulk viscosity also has significant effects. In fact, if the bulk viscosity peaks near the crossover temperature $T_c$ to the degree indicated by recent lattice calculations in QCD without quarks, it can make the fluid cavitate --- falling apart into droplets.  
It is interesting to see a hydrodynamic calculation predicting its own breakdown, via cavitation, at the temperatures where hadronization is thought to occur in ultrarelativistic heavy ion collisions.
}

\preprint{MIT-CTP-4059}

\begin{document}
\def\vev#1{\langle#1\rangle}
\def\ov{\over}
\def\le{\left}
\def\ri{\right}
\def\ha{{1\over 2}}
\def\lam{{\lambda}}
\def\Lam{{\Lambda}}
\def\al{{\alpha}}
\def\ket#1{|#1\rangle}
\def\bra#1{\langle#1|}
\def\vev#1{\langle#1\rangle}
\def\det{{\rm det}}
\def\tr{{\rm tr}}
\def\Tr{{\rm Tr}}
\def\NN{{\cal N}}
\def\th{{\theta}}

\def\Om{{\Omega}}
\def \th{{\theta}}

\def \lam {\lambda}
\def \om {\omega}
\def \ra {\rightarrow}
\def \ga {\gamma}
\def\sig{{\sigma}}
\def\ep{{\epsilon}}
\def\apr{{\alpha'}}
\newcommand{\p}{\partial}
\def\LL{{\cal L}}
\def\HH{{\cal H}}
\def\GG{{\cal G}}
\def\TT{{\cal T}}
\def\CC{{\cal C}}
\def\OO{{\cal O}}
\def\PP{{\cal P}}
\def\tir{{\tilde r}}

\newcommand{\bea}{\begin{eqnarray}}
\newcommand{\eea}{\end{eqnarray}}
\newcommand{\nn}{\nonumber\\}

\section{Introduction}

In recent years, the comparison between data from experiments at the Relativistic Heavy Ion Collider (RHIC) at Brookhaven National Laboratory on the transverse expansion of the matter produced in ultrarelativistic nucleus-nucleus collisions with nonzero impact parameters~\cite{RHIC,RHICv2}
and calculations done using second-order relativistic viscous hydrodynamics~\cite{Baier:2006gy,Romatschke:2007mq,Song:2007fn,Dusling:2007gi,Song:2007ux,Luzum:2008cw,Song:2008si,Molnar:2008xj,Song:2008hj,Luzum:2009sb,Song:2009je,Heinz:2009xj} 
have strengthened the case that the quark-gluon plasma in QCD at temperatures above, but not too far above, the crossover from a hadron gas is a strongly coupled liquid. These comparisons indicate a shear viscosity to entropy density ratio $\eta/s$ that is within a factor of a few of $1/4\pi$~\cite{Teaney:2003kp,Heinz:2009xj}, which is the value of this ratio in any gauge theory with a dual gravity description in the limit of infinite coupling and infinitely 
many colors~\cite{Policastro:2001yc,Kovtun:2003wp,Buchel:2003tz}.     The degree of success of the hydrodynamic description of these collisions, which turn out to be creating exploding droplets of a fluid that is closer to the ideal liquid limit than water is by about two orders of magnitude --- and water {\it is} the liquid that hydrodynamics is named after --- have refocused attention on the question of when a hydrodynamic description applies and when it breaks down.  The aspect of this question that has drawn most attention is ``Before what early time in the collision is a hydrodynamic description invalid?"  This is the question of how, and how quickly, approximate local thermal equilibrium is attained.   We shall focus, instead, on the complementary question ``Assuming that a hydrodynamic description is valid starting at some early time, after what late time does this hydrodynamic description break down?"   

Ultimately, heavy ion collisions produce thousands of hadrons flying outwards toward the detector. The question of how, and when, a hydrodynamic description breaks down and how one then ends up with a cloud of hadrons flying apart is the question of how, and when, ``freezeout'' occurs.
In calculations, freezeout is typically handled in one of two ways.  One option is to choose (by hand, or via fitting the output of the calculation to data) a freezeout temperature well below the crossover temperature, and at the surface in space-time where the fluid in the hydrodynamic calculation cools to this temperature apply the Cooper-Frye prescription~\cite{Cooper:1974mv} for mapping hydrodynamic volume elements directly onto a  phase space distribution of noninteracting hadrons.  The second option is to choose a temperature just below the crossover at which to map the hydrodynamic description onto a hadronic transport code, and then let this code describe hadron-hadron interactions until the hadrons later freeze out~\cite{Bass:2000ib}.    The second option is an improvement on the first, but it would be even better if the hydrodynamic description itself would tell us when it breaks down.  This is impossible within the framework of ideal (zero viscosity) hydrodynamics. An ideal hydrodynamic evolution is fully specified by its initial conditions and by a thermodynamic equation of state $p(\varepsilon)$ giving the pressure in terms of the energy density.  Given $p(\varepsilon)$, an ideal hydrodynamics code will blithely let the expanding fluid evolve until $\varepsilon\rightarrow 0$, without ever giving any hint that in reality it has gone far beyond the epoch when hydrodynamics is actually a good approximation.  Thus, one must either add a freezeout prescription by hand, or choose by hand to switch from a hydrodynamic to a transport description of the expanding fluid.  Once viscous effects are included in the hydrodynamic description, however, it {\it is} possible for the hydrodynamic equations to tell us ``from within'' when they break down.  

There are many ways in which a hydrodynamic evolution of an expanding fluid may break down, but we shall focus only on one: cavitation.   In an ordinary flowing liquid, cavitation is the formation of bubbles of vapor in regions where the pressure of the liquid drops below its vapor pressure~\cite{CavitationBook}.  Cavitation is well studied, both experimentally and theoretically. It can occur on the trailing edges of pump and propeller blades; in this context, the challenge is often to design the blade so as to avoid cavitation since when the bubbles produced by cavitation later collapse they make shock waves that can damage the blades.   Cavitation is also used in medicine --- one of the treatments of kidney stones involves destroying them via cavitation induced by an ultrasonic pulse.  Returning to our context, we shall ask whether the pressure in the expanding droplet of quark-gluon plasma can become negative, which means that it goes below the pressure of the vacuum --- our analogue of the vapor phase.   This is impossible in ideal hydrodynamics: the thermodynamic $p(\varepsilon)$ is positive.  But, in an expanding fluid in the presence of shear and bulk viscosity, shear and bulk stresses make an additional contribution to the space-space components of the stress energy tensor $T^{\mu\nu}$ and these shear and bulk stresses can drive the pressure --- which we shall denote $P$ --- negative.  In particular,  bulk viscosity is, in essence, a drop in the pressure of an expanding fluid relative to the pressure of an equilibrium fluid at the same energy density, and this decrease in the pressure will play a key role in our considerations. We shall only analyze boost-invariant $1+1$-dimensional expansion of a $3+1$-dimensional fluid~\cite{Bjorken:1982qr}, 
meaning that the fluid is only expanding in the $z$-direction.    We shall ask whether, and when, the shear and bulk stresses are sufficient to drive the longitudinal pressure (which we shall denote $P_\xi$) negative.   If $P_\xi$ goes negative in the hydrodynamic equations that describe boost-invariant expansion, what will happen when the fluid expands to the point where $P_\xi=0$?  At this point, instead of continuing to expand, dilute, and cool in a spatially uniform boost-invariant fashion as before, the fluid will break apart into fragments, separated from each other by vacuum.  Vacuum regions are the analogue of the vapor bubbles in conventional cavitation, and their pressure is zero.  So, when $P_\xi=0$ regions of fluid can stably  coexist with regions of vacuuum.  The fragments of fluid formed via cavitation will then fly apart, separating from each other, and will subsequently hadronize.   Our boost invariant calculation will only allow us to gauge whether and when $P_\xi$ goes negative: once the fluid cavitates, it is no longer boost invariant, and so our calculation will not be able to describe the subsequent dynamics.  Describing the size distribution of the fragments that results from cavitation would first of all require including transverse expansion in the hydrodynamic description, and would second of all require estimating the surface tension associated with the interface between fragments of quark-gluon plasma fluid and vacuum.  If this surface tension is large, large fragments will result; if it is small, smaller ones will be favored.

We find that as long as $\eta/s$ is in the vicinity of $1/4\pi$ the shear stress alone is not enough to trigger cavitation.\footnote{Well below the crossover temperature $T_c$, in the hadron gas, $\eta/s$ rises significantly~\cite{Prakash:1993kd,Demir:2009qi}.  If the bulk stress does not do so earlier, the shear stress could trigger cavitation at some temperature well below $T_c$.}  However, there is now evidence from a variety of directions~\cite{Paech:2006st,Kharzeev:2007wb,Meyer:2007dy,Karsch:2007jc,Sasaki:2008fg,Meyer:2009jp} that the bulk viscosity $\zeta$ is large --- $\zeta/s \sim {\cal O}(1) \gg \eta/s$ --- in a narrow range of temperatures around $T_c$.  We shall quantify how large this peak in $\zeta/s$ must be if it is to trigger cavitation when the expanding fluid has cooled to a temperature near $T_c$. We find that as long as this peak is between 1/4 and 4 times as wide as suggested by current lattice calculations (whose uncertainties we shall discuss), cavitation will occur if the peak is higher than a threshold height that lies between 1/2 and 1/4 that suggested by the lattice calculations.

Our paper is organized as follows.  In Section 2 we set up the hydrodynamic equations that describe the boost invariant $1+1$-dimensional expansion of a $3+1$-dimensional fluid,
working to second order in derivatives of the velocity field~\cite{Muronga:2001zk,Teaney:2003kp,Muronga:2003ta,Muronga:2004sf,Heinz:2005bw,Baier:2006um,Baier:2007ix,Luzum:2008cw,Fries:2008ts,Martinez:2009mf}.  After setting the full problem up, in the remainder of Section 2 we set the bulk viscosity to zero.  We describe how we specify the equation of state (which arises at zeroth order), shear viscosity (first order), and the various coefficients that arise at second order, as well as the initial conditions.  We then show results and explore their sensitivity to the second order coefficients and to the shear viscosity.  We find much greater sensitivity to the shear viscosity, indicating that, as other authors have found previously, we are using the second order equations in a regime in which the second order effects are much smaller than the first order effects.  We turn bulk viscosity on 
in Section 3.    We first describe how we parametrize $\zeta/s$ and the one new second order coefficient that at a minimum must  be introduced and in so doing mention some of the uncertainties in our current knowledge of both.  We then present our results.  In Section 4 we speculate about their implications.  The possibility that bulk viscosity could cause the expanding fluid to break apart into fragments has been discussed previously by Torrieri, Tomasik and Mishustin~\cite{Torrieri:2007fb} using the formalism of Ref.~\cite{Kouno:1989ps}.  We close by sketching several facets of the observed phenomenology of heavy ion collisions that could indicate that freezeout is preceded by cavitation, some previously highlighted in Refs.~\cite{Torrieri:2007fb} and some not, some coming from recent analyses of data and some of long standing.\footnote{Explorations of possible consequences of bulk viscosity in the phenomenology of heavy ion collisions that are not related to cavitation can be found in Refs.~\cite{Fries:2008ts,Denicol:2009am,Monnai:2009ad,Song:2009je}.} Because we are neglecting transverse expansion throughout we will not be able to make quantitative contact with data. But, our results motivate the importance of including the peak in the bulk viscosity near $T_c$ in hydrodynamic calculations that do include transverse expansion, and the importance of looking for cavitation in these more realistic calculations.

\section{Second order relativistic hydrodynamics for a boost-invariant $1+1$-dimensional expansion}

\subsection{Setup}

The energy momentum tensor for relativistic hydrodynamics can be written as 
\begin{equation}
T^{\mu\nu}=\varepsilon \, u^\mu\,u^\nu\, - p\, \Delta^{\mu\nu} + \Pi^{\mu\nu}
\label{Tmunu}
\end{equation}
where $\varepsilon$ and $p$ are the fluid energy density and pressure, $u^\mu$ is the fluid four-velocity, normalized such that $u_\mu u^\mu=1$, $\Pi^{\mu\nu}$ is the viscous tensor, satisfying 
$u_\mu\Pi^{\mu\nu}=0$, and where the projector
\begin{equation}
\Delta^{\mu\nu}\equiv g^{\mu\nu}-u^{\mu}u^\nu
\end{equation}
is also orthogonal to $u^{\mu}$.  Hydrodynamics is the effective theory describing the long-wavelength dynamics of the energy density and the fluid velocity.  Its evolution equations describe the conservation of energy and momentum, and are given by 
\begin{equation}
D_\mu T^{\mu\nu}=0
\label{DmuTmunu}
\end{equation}
where $D_\mu$ is the geometric covariant derivative.  We shall only be interested in hydrodynamics in flat spacetime, but it will be convenient to use curvilinear coordinates to describe boost invariant expansion and we therefore keep the geometric notation.   With $T^{\mu\nu}$ as in (\ref{Tmunu}), the evolution equations take the form
\begin{eqnarray}
\left(\varepsilon + p\right) D u^\mu &=& \nabla^\mu p - \Delta^\mu_\alpha D_\beta \Pi^{\alpha\beta}\ ,\nonumber\\
D \varepsilon &=& -\left(\varepsilon + p\right) \nabla_\mu u^\mu+\frac{1}{2} \Pi^{\mu\nu}\nabla_{\langle\nu}u_{\nu\rangle}\ ,
\end{eqnarray}
where $D\equiv u^\mu D_\mu$ is the comoving time derivative in the fluid rest frame, $\nabla^\mu \equiv \Delta^{\mu\nu}D_\nu$ is the spatial derivative in the fluid rest frame, and the brackets $\langle\ldots\rangle$ denote the combination that is symmetric, traceless, and orthogonal to the fluid velocity, namely
\begin{equation}
A_{\langle\mu}B_{\nu\rangle}\equiv \left( \Delta^\alpha_\mu \Delta^\beta_\nu + \Delta^\alpha_\nu \Delta^\beta_\mu -\frac{2}{3}\Delta^{\alpha\beta}\Delta_{\mu\nu}\right)A_\alpha B_\beta\ .
\end{equation}
In general, $\Pi^{\mu\nu}$ includes the physics of shear viscosity, bulk viscosity and thermal conductivity.  Thermal conductivity is only relevant if there is a nonzero density of some species with a conserved particle number, and we shall work at zero baryon chemical potential throughout.  In a conformal fluid, the bulk viscosity vanishes and, including terms up to second order in gradients, 
$\Pi^{\mu\nu}$  satisfies~\cite{Baier:2007ix,Bhattacharyya:2008jc} 
\begin{eqnarray}
\Pi^{\mu\nu}&=&\eta \nabla^{\langle\mu}u^{\nu\rangle} - \tau_{\Pi}^\eta\left[\Delta^\mu_\alpha \Delta^\nu_\beta D \Pi^{\alpha\beta} + \frac{4}{3} \Pi^{\mu\nu} \nabla_\alpha u^\alpha \right]\nonumber\\
&~&\qquad -\frac{\lambda_1}{2\eta^2}\Pi^{\langle\mu}_\alpha\Pi^{\nu\rangle\alpha}
+\frac{\lambda_2}{2\eta}\Pi^{\langle\mu}_\alpha \omega^{\nu\rangle\alpha}
-\frac{\lambda_3}{2}\omega^{\langle\mu}_\alpha \omega^{\nu\rangle\alpha}\ ,
\label{Pimunu}
\end{eqnarray}
where $\omega_{\mu\nu}\equiv -\frac{1}{2}\left(\nabla_\mu u_\nu  - \nabla_\nu u_\mu\right)$ is the fluid vorticity, where the shear viscosity $\eta$ is the only property of the fluid that enters at first order in gradients, and where $\tau_\Pi^\eta$, $\lambda_1$, $\lambda_2$ and $\lambda_3$ are the four properties of the fluid that arise at second order.  Obtaining a closed set of evolution equations requires specifying the equation of state $p(\varepsilon)$ and specifying $\eta$, $\tau_\Pi$ and $\lambda_{1,2,3}$ in terms of $\varepsilon$.   Equivalently, $p$, $\varepsilon$, $\eta$, $\tau_\Pi$ and $\lambda_{1,2,3}$ can all be specified in terms of the temperature $T$.    It is sometimes also convenient to introduce the entropy density 
\begin{equation}
s=\frac{p+\varepsilon}{T}\ .
\label{EntropyDefn}
\end{equation}
Conformality determines the equation of state $p=\varepsilon/3$ and implies that $p=\frac{\varepsilon}{3}\propto T^4$, $\eta\propto T^3$, 
$\tau_\Pi^\eta\propto T^{-1}$, and $\lambda_{1,2,3}\propto T^2$, but conformality alone does not determine any of  the dimensionless proportionality constants other than the one in the equation of state.

If we relax the assumption of conformality (while continuing to assume throughout that there is no net baryon density) the only new term that arises on the right-hand side of (\ref{Pimunu}) that is 
first-order in derivatives is  $-\zeta (\nabla_\alpha u^\alpha) \Delta^{\mu\nu}$, where $\zeta$ is the bulk viscosity.   At second order, many new terms arise~\cite{Romatschke:2009kr}. As we shall discuss below, it is a standard simplifying assumption to write only the term $+\zeta \tau_\Pi^\zeta D(\nabla_\alpha u^\alpha)\Delta^{\mu\nu}$, where $\tau_\Pi^\zeta$ is a new second order coefficient whose role we discuss below.

We shall only consider solutions to the $3+1$-dimensional hydrodynamic equations in which no quantity depends on the transverse spatial directions $x$ and $y$ (which in particular means no vorticity) and in which the expansion in the $z$-direction is boost invariant~\cite{Bjorken:1982qr}.  This makes it convenient to change variables from $(t,z)$ to $(\tau,\xi)$ where $\tau\equiv\sqrt{t^2-z^2}$ is the proper time and $\xi\equiv {\rm ArcTanh}\left(z/t\right)$ is the spacetime rapidity. These curvilinear coordinates are comoving with the fluid, meaning that $u^\tau=1$ and the spatial components of $u$ all vanish. And, in these coordinates boost invariance implies significant further simplifications:  $\Pi^{\mu\nu}$ is diagonal, and therefore so is $T^{\mu\nu}$, and the diagonal components of $T^{\mu\nu}$ depend only on $\tau$, not on $\xi$.   Upon making these simplifications, the stress energy tensor in $(\tau,x,y,\xi)$ coordinates takes the form~\cite{Muronga:2001zk,Teaney:2003kp,Muronga:2003ta,Muronga:2004sf,Heinz:2005bw,Baier:2006um,Baier:2007ix,Luzum:2008cw,Fries:2008ts,Martinez:2009mf}
\begin{equation}
T^{\mu\nu}=
\left( \begin{array}{cccc}
\varepsilon &\  0 &\  0 &\  0  \\
0 & \ p & \ 0 & \ 0 \\
0 &\ 0 & \ p &\  0 \\
0 &\  0 & \ 0 & \ p \end{array}\right) + 
\left( \begin{array}{cccc}
0 &\  0 & \ 0 &\  0  \\
0 & \ \Pi +
\frac{1}{2}\Phi  &\  0 &\  0 \\
0 &\  0 &\  \Pi + \frac{1}{2} \Phi &\  0 \\
0 &\  0 &\  0 &\  \Pi - \Phi  \end{array} \right) \ ,
\end{equation}
where the trace of $\Pi^{\alpha\beta}$ --- namely $\Pi$ --- and the traceless part of $\Pi^{\alpha\beta}$ --- namely $\Phi$ ---  denote the non-equilibrium contributions to the pressure coming from the bulk and shear stresses, respectively.\footnote{The quantity that we denote as $\Phi$ has been called $\Phi$ in some of the literature and $\Pi$ elsewhere in the literature.}   
For a fluid at rest, the pressure is isotropic and is given by $p$, which is related to the energy density by the thermodynamic equation of state $p(\varepsilon)$.  As the fluid is expanding, unless it is an ideal fluid with $\Pi^{\alpha\beta}=0$ its pressure is no longer isotropic --- the transverse and longitudinal pressure are given by
\begin{eqnarray}
P_\perp &\equiv& p + \Pi + \frac{1}{2}\Phi\label{PPerp}\\
P_\xi &\equiv& p + \Pi - \Phi\ .\label{PLong}
\end{eqnarray}
Furthermore, upon making these simplifications the second order evolution equations are
\begin{eqnarray}
\frac{\partial\varepsilon}{\partial\tau} &=& -\frac{\varepsilon+p+\Pi-\Phi}{\tau}\ ,\label{EpsEquation}\\
\tau_\Pi^\eta\, \frac{\partial \Phi}{\partial \tau} &=& \frac{4\eta}{3\tau}-\Phi-\left[ \frac{4\tau_\Pi^\eta}{3\tau}\,\Phi + \frac{\lambda_1}{2\eta^2}\,\Phi^2 \right]\ ,\label{PhiEquation}\\
\tau_\Pi^\zeta \,\frac{\partial \Pi}{\partial \tau} &=& - \frac{\zeta}{\tau} - \Pi \ .  \label{PiEquation}
\end{eqnarray}
At first order, $\Pi$ and $\Phi$ are given by their Navier-Stokes values
\begin{equation}
\Phi=\frac{4\eta}{3\tau}
\label{PhiNavierStokes}
\end{equation}
and
\begin{equation}
 \Pi=-\frac{\zeta}{\tau}\ .
 \label{PiNavierStokes}
\end{equation}
We see that if we ignore the terms in square brackets in (\ref{PhiEquation}), then the second order equations (\ref{PhiEquation}) and (\ref{PiEquation}) describe $\Phi$ and $\Pi$ relaxing towards their Navier-Stokes behavior (\ref{PhiNavierStokes}) and (\ref{PiNavierStokes}) with time constants $\tau_\Pi^\eta$ and $\tau_\Pi^\zeta$, along the lines of the Israel-Stewart approach to second order dissipative relativistic hydrodynamics~\cite{IsraelStewart}. If we ignore bulk viscosity, setting $\Pi=0$, then (\ref{EpsEquation}) and (\ref{PhiEquation}), including in particular the terms in the square brackets in (\ref{PhiEquation}), follow from conformality~\cite{Baier:2007ix,Bhattacharyya:2008jc}.  However, once we turn on bulk viscosity we are breaking conformality, and there can then be further second order terms in both (\ref{PhiEquation}) and (\ref{PiEquation})~\cite{Romatschke:2009kr}.  These equations are in this sense provisional, but it should be noted that the terms in square brackets in (\ref{PhiEquation}) become neglible at large $\tau$ and we expect the same to be the case for the missing nonconformal terms also.  

\subsection{Signs of cavitation}

Since $\Phi>0$ and $\Pi<0$ at first order, see (\ref{PhiNavierStokes}) and (\ref{PiNavierStokes}), it is reasonable to expect them to have these signs in solutions to the second order equations also. We then see from (\ref{PPerp}) and (\ref{PLong})
that if either $\zeta$ or $\eta$ is large enough, the longitudinal 
pressure $P_\xi$ can be driven negative, and if $\zeta$ is large enough, the transverse pressure $P_\perp$ can also be driven negative.   
We shall see in Section 3 that if $\zeta$ rises high enough at temperatures in the vicinity of the crossover from quark-gluon plasma to hadron gas, the resulting bulk stress $\Pi$  does drive $P_\xi$ negative, indicating cavitation.

\subsection{Equation of state, shear viscosity, $\tau_\Pi^\eta$ and $\lambda_1$}

We see that in order for the evolution equations (\ref{EpsEquation}),  (\ref{PhiEquation}) and (\ref{PiEquation})  to be closed we need the equation of state $p(\varepsilon)$ and expressions relating $\eta$, $\zeta$, $\tau_\Pi^\eta$, $\tau_\Pi^\zeta$ and $\lambda_1$ to $\varepsilon$. 
For the remainder of Section 2, we shall set $\zeta=0$ and therefore $\Pi=0$, deferring our analysis of the effects of bulk viscosity to Section 3.  We then need $p$, $\eta$, $\tau_\Pi^\eta$ and $\lambda_1$ only.

We need an equation of state $p(\varepsilon)$ that describes the quark-gluon plasma phase as well as the crossover to a hadron gas.  Lattice quantum field theory is well-suited to the calculation of thermodynamic quantities at zero baryon chemical potential, and so there are many lattice calculations of $p(\varepsilon)$ in QCD that we could employ.  We shall take $p(\varepsilon)$ from Ref.~\cite{Bazavov:2009zn}, both because it is an example of the state of the art and because these authors have provided a parametrization of their results that is easy to use.   They parametrize their results for the trace anomaly using the functional form
\begin{equation}
\frac{\varepsilon-3p}{T^4} = \left(1-\frac{1}{\left[1+\exp\left(\frac{T-c_1}{c_2}\right)\right]^2}\right)\left(\frac{d_2}{T^2}+\frac{d_4}{T^4}\right)\ ,
\label{TraceAnomaly}
\end{equation}
and give values with error bars for the coefficients $d_2$, $d_4$, $c_1$ and $c_2$ for calculations done with two different lattice actions, with and without combining these calculations with hadron resonance gas calculations valid at lower temperatures.  We shall use the central values of their results obtained from combining lattice calculations done with the p4 action and hadron resonance gas calculations: $d_2= 0.24$~GeV$^2$, $d_4=0.0054$~GeV$^4$, $c_1=0.2073$~GeV, and $c_2=0.0172$~GeV.  These authors find a crossover between hadron gas and quark-gluon plasma occurring in a temperature regime $180$~MeV$\lesssim T\lesssim 200$~MeV. In Section 3 when we need to specify a value of the crossover temperature $T_c$ we shall use $T_c=190$~MeV, in order to be consistent with the equation of state that we employ throughout.
The pressure is related to the trace anomaly by
\begin{equation}
\frac{p(T)}{T^4} - \frac{p(T_0)}{T_0^4} = \int_{T_0}^T dT' \,\frac{\varepsilon-3p}{T'^5}\ ,
\label{LatticeEOS}
\end{equation}
and the results of Ref.~\cite{Bazavov:2009zn} are obtained by choosing $T_0=50$~MeV and $p(T_0)=0$.  We shall only work at $T>100$~MeV, where there is no effect of these choices.
Knowing $(\varepsilon-3p)$ and $p$ as functions of $T$, we know $\varepsilon$ as a function of $T$ also, as well as the entropy density (\ref{EntropyDefn}).  And, from $p(T)$ and $\varepsilon(T)$ we have the equation of state $p(\varepsilon)$.  We shall use the same equation of state throughout this paper, focussing on the effects of varying other quantities.

Next, we turn to the shear viscosity $\eta$.  We shall use 
\begin{equation}
\frac{\eta}{s}=\frac{1}{4\pi}
\label{KSS}
\end{equation}
as a baseline value, and we shall explore the effects of varying $\eta/s$.  The relationship (\ref{KSS}) holds for the plasma phase 
of any gauge theory 
that has a dual gravity description, in the limit of large numbers of colors and infinitely strong coupling~\cite{Policastro:2001yc,Kovtun:2003wp,Buchel:2003tz}.  
Even though much larger values of $\eta/s$ (of order 1 and larger) are expected both in the hadron gas found well below $T_c$~\cite{Prakash:1993kd,Demir:2009qi} and in the weakly coupled quark-gluon 
plasma found far above $T_c$~\cite{Arnold:2003zc}, the baseline (\ref{KSS}) is seen as a reasonable starting point for the analysis of quark-gluon plasma in the regime being explored by RHIC collisions, say around $T\sim 1.5 T_c$.  
Lattice QCD calculations, to date in a gluon plasma without quarks, 
indicate values of $\eta/s$  only a few times (\ref{KSS})~\cite{Meyer:2007ic,Meyer:2009jp}: at $T=1.58 T_c$ Meyer finds $(\eta+\frac{3}{4}\zeta)/s=0.20$ with a statistical error of $\pm.03$~\cite{Meyer:2009jp}. (At this temperature, $\zeta$ is small compared to $\eta$.)  
Comparison between second order relativistic viscous hydrodynamic calculations that include 
transverse expansion~\cite{Baier:2006gy,Romatschke:2007mq,Song:2007fn,Dusling:2007gi,Song:2007ux,Luzum:2008cw,Song:2008si,Molnar:2008xj,Song:2008hj,Luzum:2009sb,Song:2009je,Heinz:2009xj} and data from RHIC~\cite{RHIC,RHICv2} on the azimuthal anisotropy of ultrarelativistic heavy ion collisions that have a significant impact parameter indicate that in these collisions approximate local thermal equilibrium is  attained rapidly and $\eta/s$ is small, apparently $\lesssim 0.2$ and conservatively $\lesssim 0.5$~\cite{Teaney:2003kp,Heinz:2009xj}.  Given that we analyze longitudinal expansion only, we can have nothing to say about the extraction of information about $\eta/s$ from these data.   But, we shall confirm that  $1+1$-dimensional boost invariant expansion is modified significantly 
as we vary $\eta/s$ between 0 and $2/4\pi$.  

Less is known about the values of the second-order coefficients $\tau_\Pi^\eta$ and $\lambda_1$.  We shall take as a baseline
\begin{equation}
\tau_\Pi^\eta = \frac{2-\log 2}{2\pi T}
\label{tauPiBaseline}
\end{equation}
and
\begin{equation}
\lambda_1 = \frac{\eta}{2\pi T} =  \frac{s}{8\pi^2 T}\ ,
\label{lambda1Baseline}
\end{equation}
which are their values in the plasma of ${\cal N}=4$ supersymmetric Yang-Mills 
theory~\cite{Baier:2007ix,Bhattacharyya:2008jc,Natsuume:2007ty},  the simplest and best studied example of a strongly coupled plasma with a dual gravity 
description.    Meyer finds $2\pi T\,\tau_\Pi^\eta = 3.1\pm0.3$ at $T=1.58 T_c$ in lattice calculations of QCD without quarks~\cite{Meyer:2009jp}, within a factor of a few of (\ref{tauPiBaseline}).\footnote{In kinetic theory, $\tau_\Pi^\eta=(5.0~{\rm to}~5.9 )\eta/(Ts)$~\cite{York:2008rr,Baier:2006um,Huovinen:2008te}, with the prefactor depending on the value of the coupling constant.  Kinetic theory is not quantitatively valid if $\eta/s=1/4\pi$, but applying it anyway gives $\tau_\Pi^\eta$ in agreement with Meyer's lattice result. In kinetic theory, $\lambda_1=(4.1~{\rm to}~5.2) \eta^2/(Ts)$~\cite{York:2008rr}, which with $\eta/s=1/4\pi$ would give a value of $\lambda_1$ within a factor of two of (\ref{lambda1Baseline}).} We shall find that the effects of varying $\tau_\Pi^\eta$ and $\lambda_1$ by large factors are small, indicating that the hydrodynamic calculations are being done in a regime in which second order effects are small compared to first order effects.

\subsection{Baseline results}

\begin{figure}[t]
\vskip-0.2in
\hskip-0.2in
\includegraphics[width=7.5cm,angle=270]{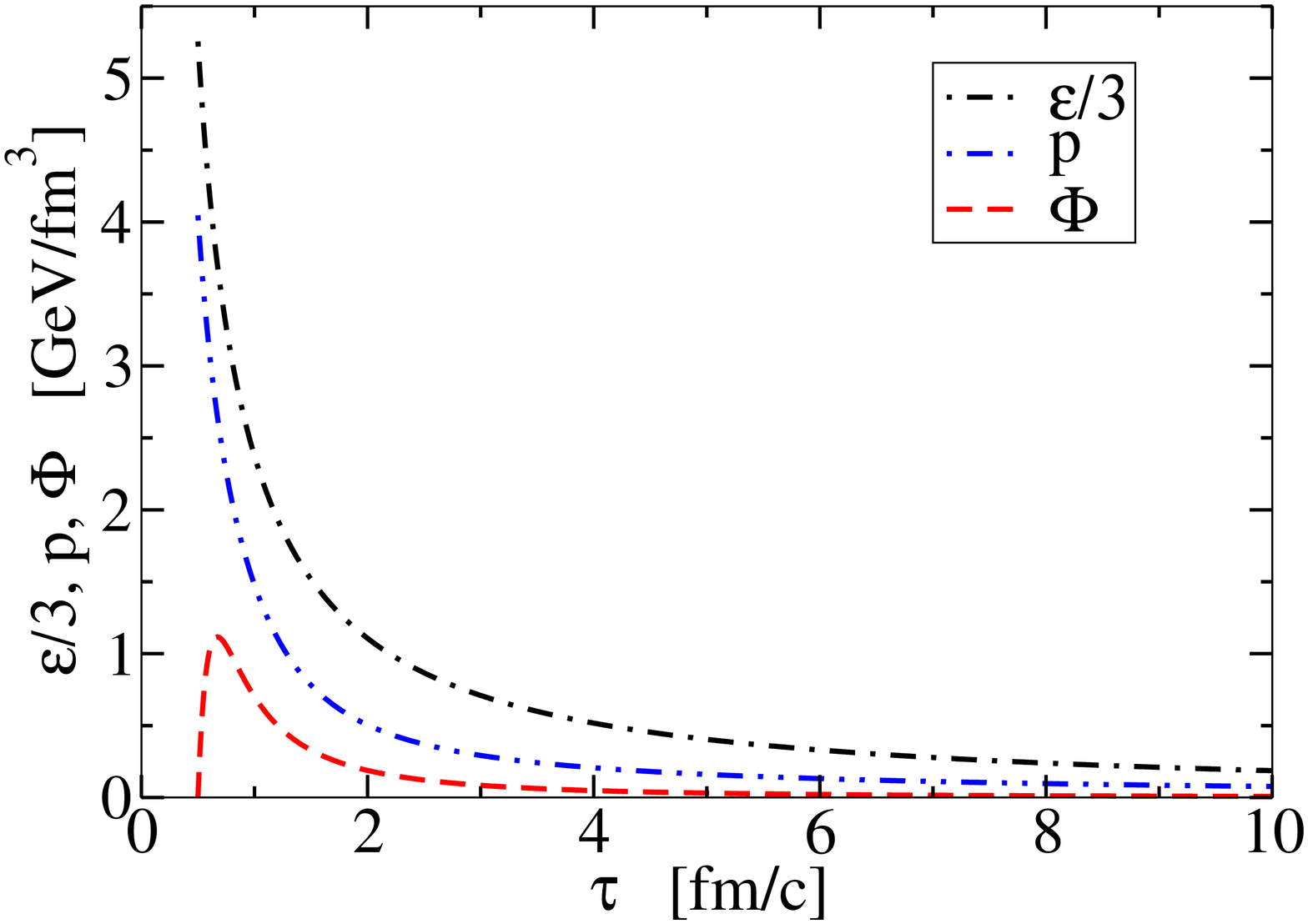}
\hskip-0.1in
\includegraphics[width=7.5cm,angle=270]{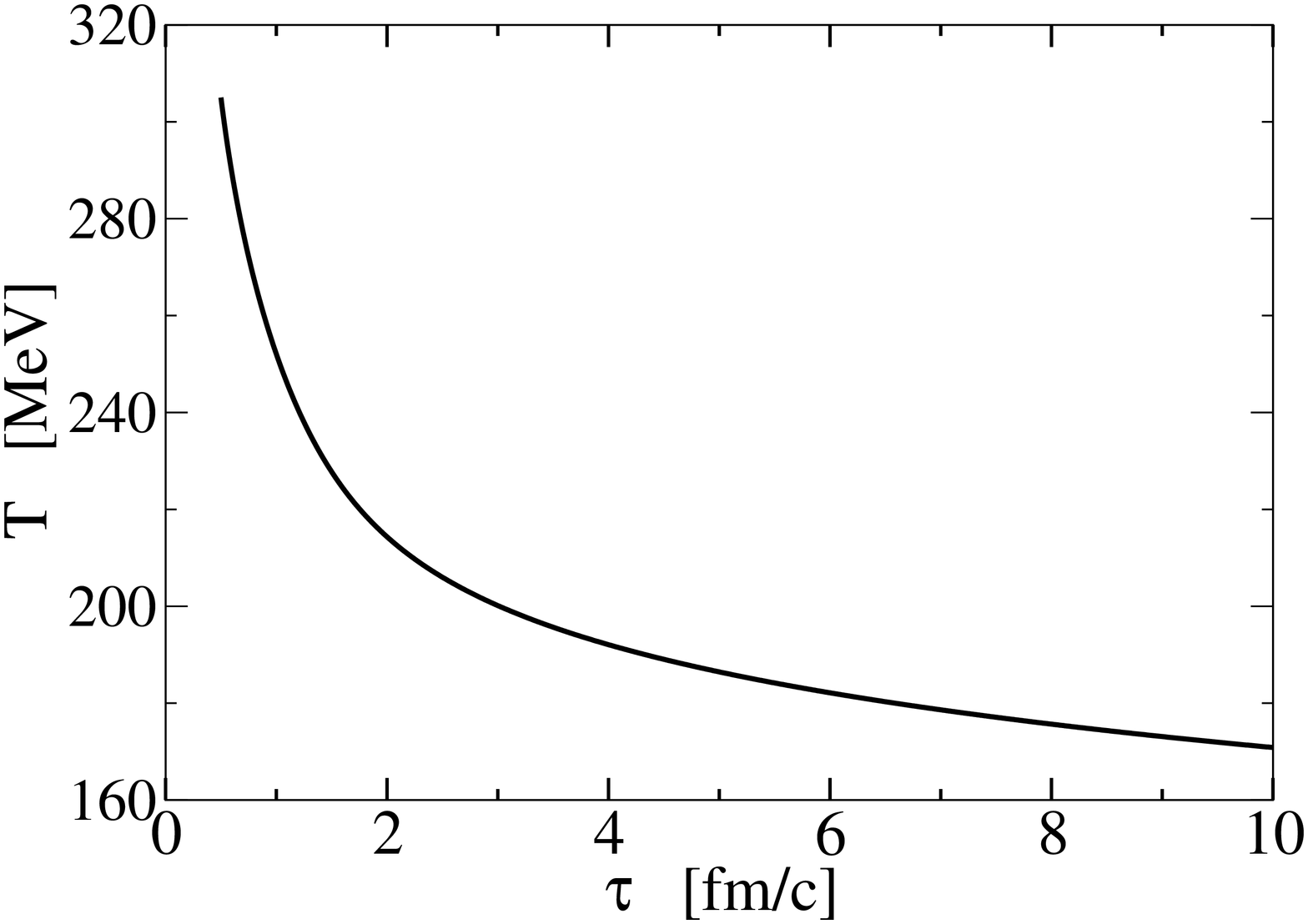}
\vskip-0.05in
\caption{
Evolution of the energy density $\varepsilon$ (plotted as $\varepsilon/3$), 
pressure $p$, and shear stress $\Phi$ (left panel) and temperature $T$ (right panel) as functions of proper time $\tau$.  The equation of state is taken from lattice calculations of QCD thermodynamics as described in Section 2.3 and the shear viscosity $\eta$ and $\tau_\Pi^\eta$ and $\lambda_1$ have all been set to the baseline values described in Section 2.3.  The bulk viscosity is zero. The evolution starts at $\tau=0.5$~fm$/c$  with the energy density $\varepsilon$ being that at $T=305$~MeV. The evolution starts with $\Phi=0$.
}
\label{Fig:baseline}
\end{figure}

In Fig.~\ref{Fig:baseline} we show an example of a solution to the evolution equations (\ref{EpsEquation}) and (\ref{PhiEquation}) with vanishing bulk viscosity.  The pressure $p$ and temperature $T$ are related to the energy density $\varepsilon$ via the lattice calculations of QCD thermodynamics from Ref.~\cite{Bazavov:2009zn} that we have described in Section 2.3.  The shear viscosity $\eta$ and the second-order coefficients $\tau_\Pi^\eta$ and $\lambda_1$ have been set to their baseline values (\ref{KSS}), (\ref{tauPiBaseline}) and (\ref{lambda1Baseline}).  

We have initialized the evolution in Fig.~\ref{Fig:baseline} at a time $\tau=0.5$~fm$/c$, a reasonable choice given that the RHIC data on the anisotropic expansion in heavy ion collisions at nonzero impact parameter can only be understood if a hydrodynamic description is already relevant earlier 
than 1~fm$/c$ after the collision~\cite{Kolb:2003dz}.  We have also chosen a value for the initial energy density that is reasonable for collisions at the top RHIC energy.    
At time $\tau=1$~fm$/c$ in the evolution of Fig.~\ref{Fig:baseline}, the energy density is 
$\varepsilon=7.12$~GeV$/$fm$^3$, which is consistent with estimates of the energy density at this time that have been made using data on the final state energy and entropy~\cite{RHIC,Muller:2005en}. 
(Using the lattice results for $\varepsilon(T)$, this energy density corresponds to a temperature $T=252$~MeV.) So, although it is misplaced precision to specify initial conditions with $\varepsilon=15.8$~GeV$/$fm$^3$ (and $T=305$~MeV) at $\tau=0.5$~fm$/c$, we shall make this choice throughout this paper as in Fig.~\ref{Fig:baseline} since varying these choices within a reasonable range would have no qualitative effects.  Note also that if we were doing phenomenology (which we cannot do given our $1+1$-dimensional expansion) we would want to adjust the initial state energy density as we vary other parameters (which we will do below) in order to maintain the same late time energy density.  We shall not do this, since our purpose is to explore how solutions to the evolution equations depend on parameters, and tweaking the initial conditions as we varied the parameters would for this purpose be a complication.

In Fig.~\ref{Fig:baseline}, we have chosen $\Phi=0$ at $\tau=0.5$~fm$/c$.  There is no phenomenological justification for this choice.  Instead, we find that this choice does not matter.  What we observe in the evolution is that $\Phi$ rapidly (over a timescale that is a few tenths of a fm$/c$ in Fig.~\ref{Fig:baseline} and that is controlled by $\tau_\Pi^\eta$) increases until it is close to its Navier-Stokes behavior (\ref{PhiNavierStokes}), and then follows (\ref{PhiNavierStokes}) closely during the subsequent evolution.   If instead of initializing with $\Phi=0$ we choose $\Phi$ at $\tau=0.5$~fm$/c$ to be twice its Navier-Stokes value, we find the same behavior.  And, varying the initial value of $\Phi$ over this range makes very little difference --- it changes $\varepsilon(\tau)$ by less than half as much as we shall find when we vary $\tau_\Pi^\eta$ in the next section.

We have plotted Fig.~\ref{Fig:baseline} up to a proper time of $\tau=10$~fm$/c$, when 
$T=171$~MeV, well below $T_c\sim 190$~MeV.   Extending the calculations to later times, we find $T=156$~MeV at $\tau=20$~fm$/c$, but this is not relevant because at these low temperatures, the shear viscosity of the hadron gas is much greater than the baseline value (\ref{KSS}).   

Note that in Fig.~\ref{Fig:baseline}, the shear stress $\Phi$ is less than the isotropic pressure $p$ at all times, meaning that the longitudinal pressure $P_\xi$ of (\ref{PLong}) is everywhere positive.  We shall see in Section 2.6 that this need no longer be so if larger values of $\eta/s$ are used.

\subsection{Insensitivity to $\tau_\Pi^\eta$ and $\lambda_1$}

\begin{figure}[t]
\vskip-0.2in
\hskip-0.1in
\includegraphics[width=7.5cm,angle=270]{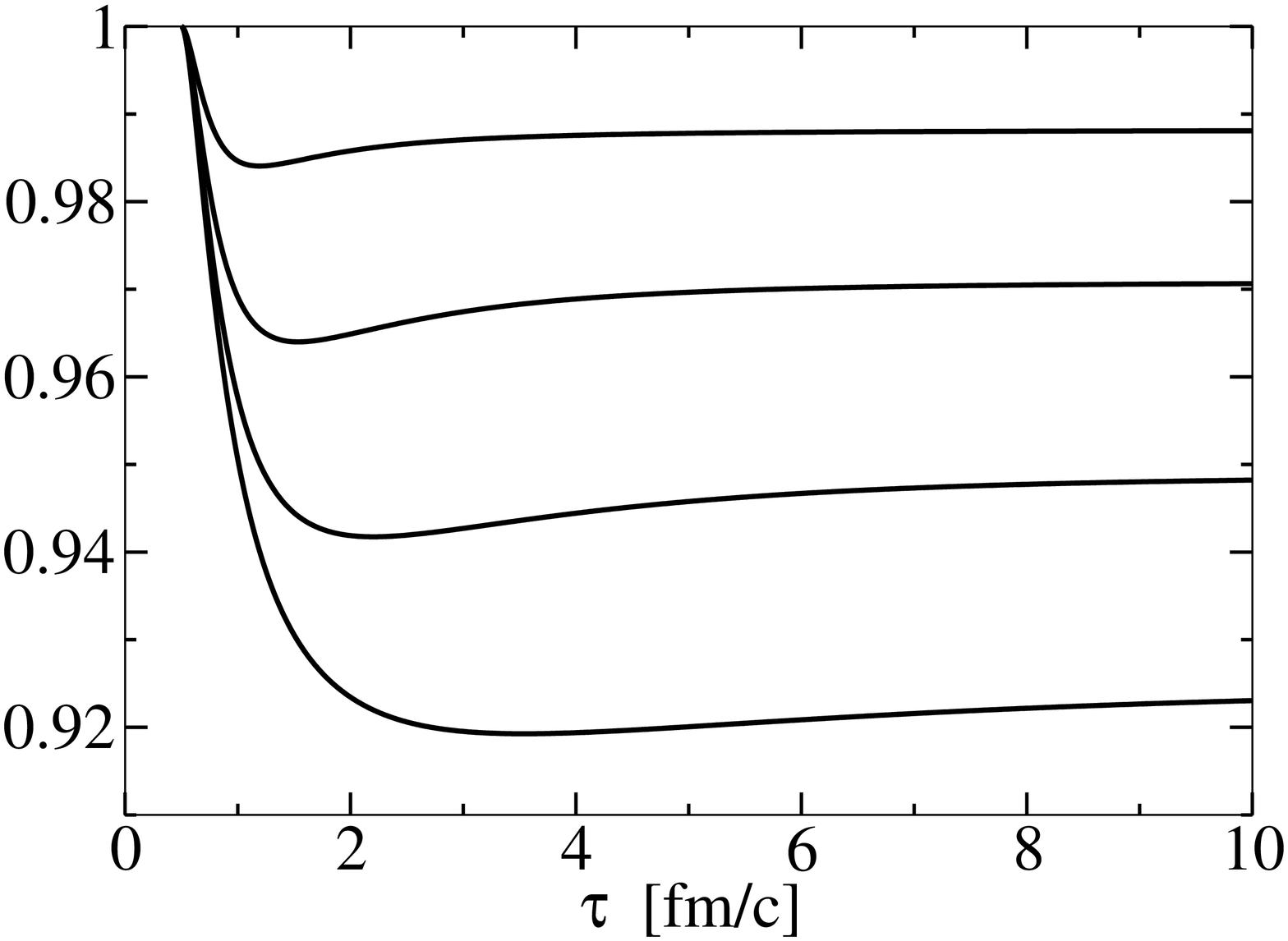}
\hskip-0.2in
\includegraphics[width=7.5cm,angle=270]{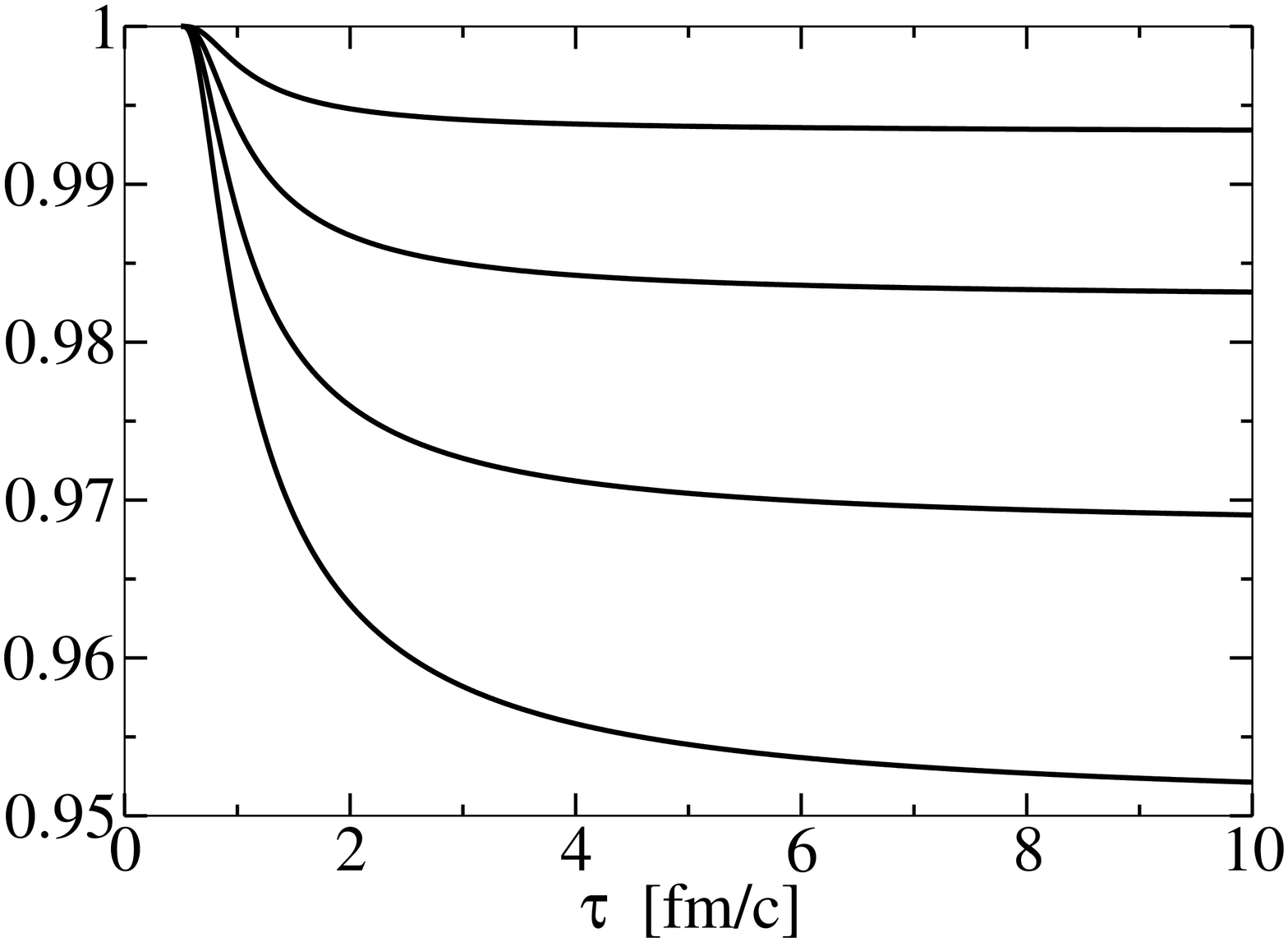}
\vskip-0.05in
\caption{
Effects of varying $\tau_\Pi^\eta$ and $\lambda_1$.  
The left panel shows the effects on the energy density $\varepsilon(\tau)$ of increasing $\tau_\Pi^\eta$. From top to bottom, the curves show $\varepsilon(\tau)$ with $\tau_\Pi^\eta$ increased by factors of 2, 4, 8 and 16 relative to its baseline  ({\protect\ref{tauPiBaseline}}) divided by $\varepsilon(\tau)$ with $\tau_\Pi^\eta$ unmodified.
The right panel shows the effects on the energy density $\varepsilon(\tau)$ of increasing $\lambda_1$ relative to its baseline ({\protect\ref{lambda1Baseline}}) by factors of (top to bottom) 2, 4, 8 and 16.     
}
\label{Fig:VaryTauPiLambda1}
\end{figure}

In Fig.~\ref{Fig:VaryTauPiLambda1} we see that both the second order coefficients $\tau_\Pi^\eta$ and $\lambda_1$ can be increased by large factors without having significant effects on the evolution of $\varepsilon(\tau)$.  Increasing $\tau_\Pi^\eta$ extends the initial period of time in the evolution when the shear stress $\Phi$ changes from its initial value 0 to approach its Navier-Stokes behavior (\ref{PhiNavierStokes}).  With $\tau_\Pi^\eta$ increased relative to its baseline value (\ref{tauPiBaseline}) by a factor of 8, these early-time transients last $1-2$~fm$/c$.  Even in this case, Fig.~\ref{Fig:VaryTauPiLambda1} shows that the modification of $\varepsilon(\tau)$ is small.  Increasing $\lambda_1$ depresses $\Phi$ relative to its Navier-Stokes behavior (\ref{PhiNavierStokes}) at all times, but the effect is small even when $\lambda_1$ is increased relative to its baseline value (\ref{lambda1Baseline}) by a factor of 16, and the effect on $\varepsilon(\tau)$ is even smaller.  To see significant effects on $\Phi$, $\lambda_1$ must be increased by factors of $\sim 100$, and even then the effect on $\varepsilon(\tau)$ is only of order 10 percent.
We conclude that we are using the evolution equations in a regime in which the effects of the second order terms are small --- smaller, we shall see, than the effects of the first order terms.  We shall therefore use the baseline values (\ref{tauPiBaseline}) and (\ref{lambda1Baseline}) exclusively in results that we shall quote throughout the following. But, we have checked that varying these parameters simultaneously with the variations of $\eta$ and $\zeta$ that we discuss below does not change any interesting conclusions.

\subsection{Sensitivity to shear viscosity}

\begin{figure}[t]
\vskip-0.2in
\hskip-0.2in
\includegraphics[width=7.5cm,angle=270]{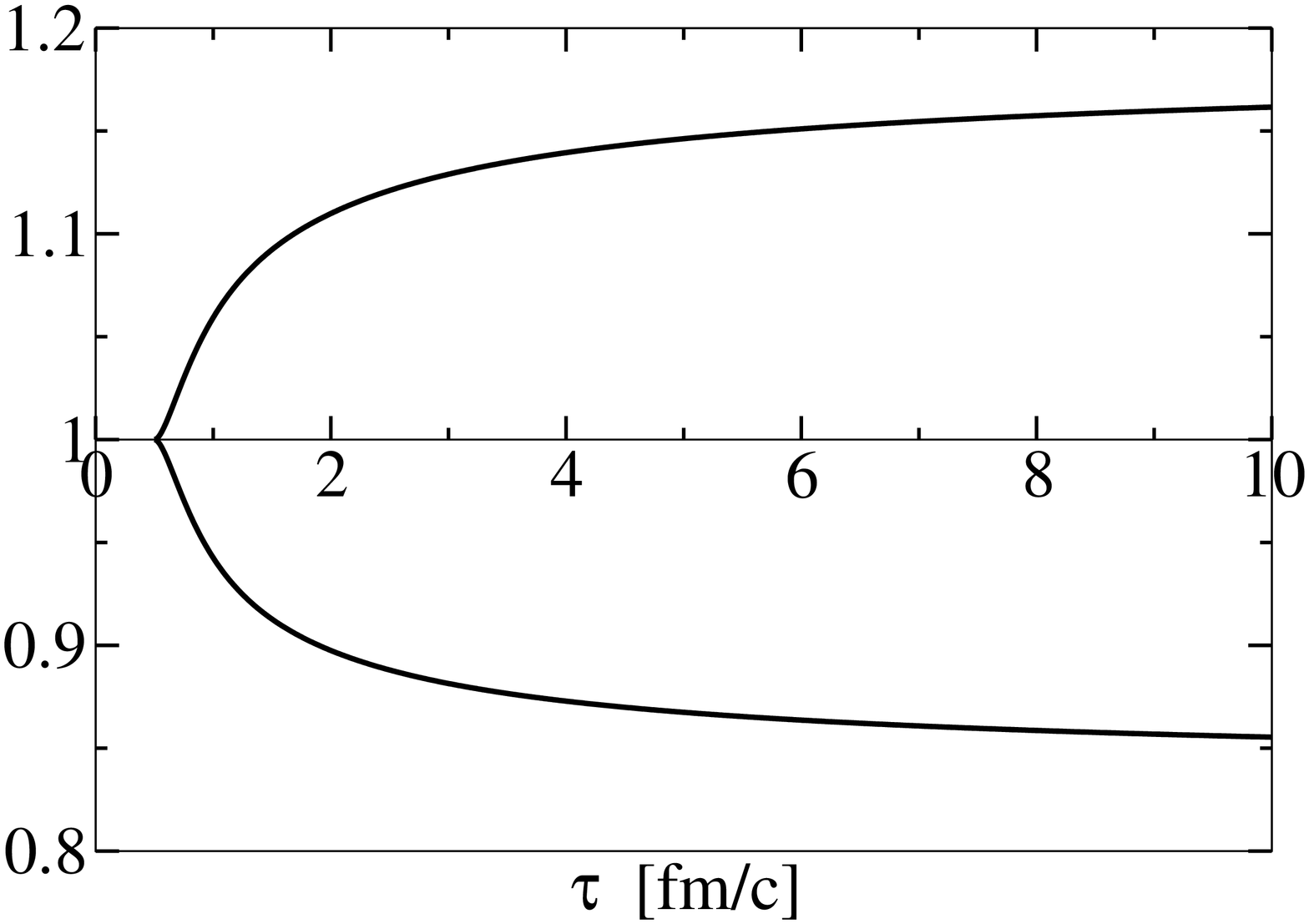}
\hskip-0.1in
\includegraphics[width=7.5cm,angle=270]{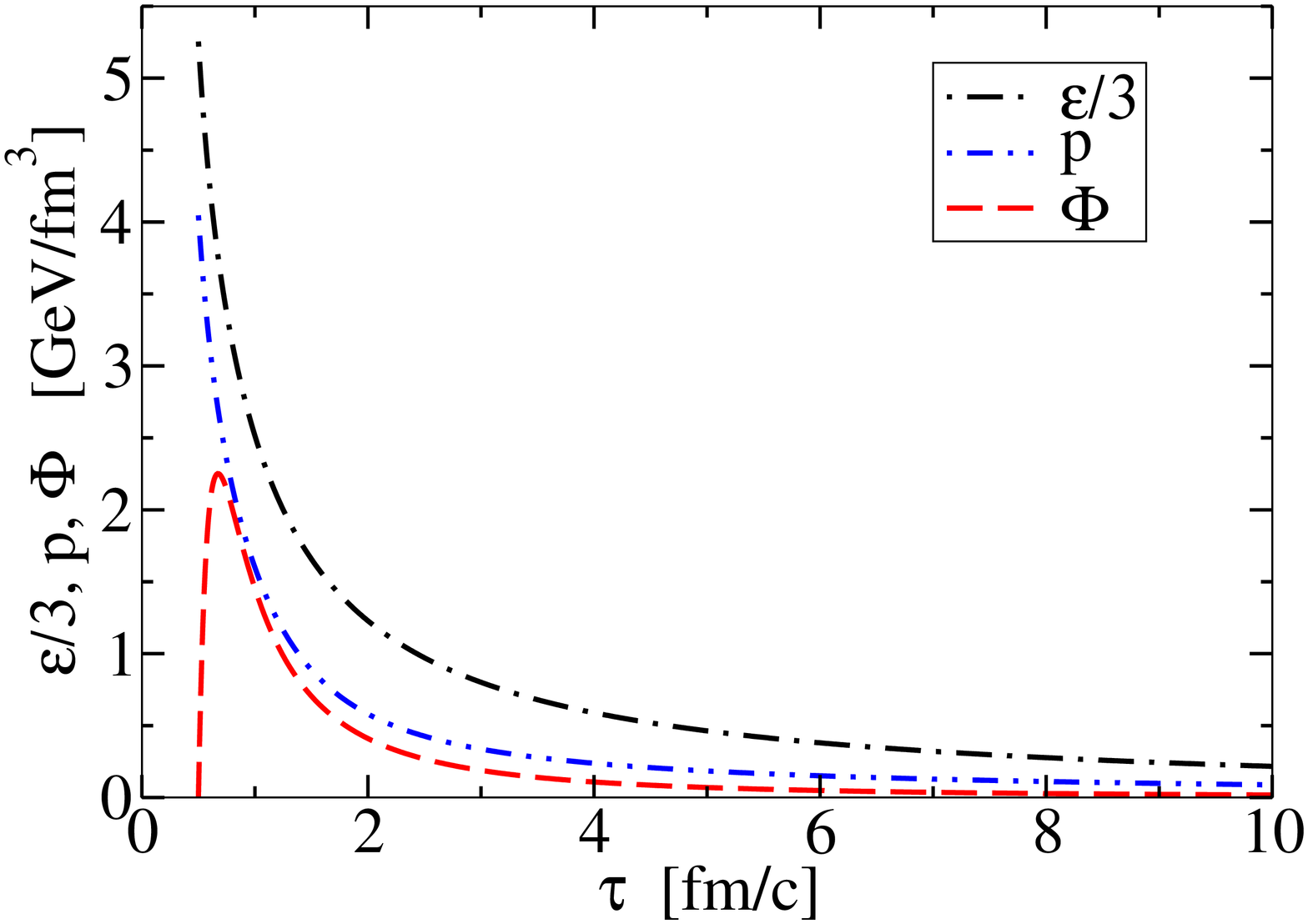}
\vskip-.05in
\caption{
Effects of varying the shear viscosity $\eta$.  The left panel shows the effects on the energy density $\varepsilon(\tau)$ of setting $\eta/s$ to $10^{-4}$ (lower curve) or $2/4\pi$ (upper curve), relative to its baseline value of $1/4\pi$. (The curves show the ratio of $\varepsilon(\tau)$ with the modified $\eta/s$ to $\varepsilon(\tau)$ with $\eta/s=1/4\pi$.)  The right panel shows $\varepsilon/3$, $p$, and the shear stress $\Phi$ as functions of $\tau$ for the case where $\eta/s=2/4\pi$.  (This panel should be compared to the left panel of Fig.~{\protect\ref{Fig:baseline}}.)  We see that the longitudinal pressure $P_\xi$, which in the absence of bulk viscosity is given by $p-\Phi$, comes close to vanishing at an early time.
}
\label{Fig:VaryEta}
\end{figure}

In Fig.~\ref{Fig:VaryEta} we see that changing the coefficient of the first order term in the evolution equations, namely the shear viscosity $\eta$, has much more significant effects than those we found in Section 2.5.  Increasing $\eta$ by a factor of two has a $\sim15$\% effect on the energy 
density $\varepsilon(\tau)$.   The fact that the evolution equations are much more sensitive to variation of $\eta$ than they are to variations of $\tau_\Pi^\eta$ or $\lambda_1$ provides qualitative support to the program~\cite{Romatschke:2007mq,Song:2007fn,Dusling:2007gi,Song:2007ux,Luzum:2008cw,Song:2008si,Molnar:2008xj,Song:2008hj,Luzum:2009sb,Song:2009je,Heinz:2009xj} 
of using comparison between hydrodynamic calculations that include anisotropic transverse expansion and data from RHIC to extract information about the value of $\eta/s$ --- this extraction should not be complicated by our lack of knowledge of the values of $\tau_\Pi^\eta$ or $\lambda$.  We will revisit this conclusion in Section 4 after considering the effects of bulk viscosity, which we have so far neglected, in Section 3.

Furthermore, we see from the right panel of Fig.~\ref{Fig:VaryEta} that, as Martinez 
and Strickland have analyzed in 
detail~\cite{Martinez:2009mf}, increasing $\eta$ makes the longitudinal pressure $P_\xi$ of (\ref{PLong}) negative at early times.    With the initial conditions that we are using, we find that $P_\xi$ is negative for some window of early $\tau$'s if $\eta/s>2.15/4\pi$.   Martinez and Strickland have analyzed how this criterion depends on the choice of initial conditions.    It is easy to see why $P_\xi<0$ must arise at early times for sufficiently large $\eta/s$.  We have seen that the shear stress rapidly rises to its Navier-Stokes value (\ref{PhiNavierStokes}). Together with $\eta\propto s$ this means that, after transient behavior at very early times, $\Phi\propto s/\tau$.  The zeroth order (ideal hydrodynamics; no shear viscosity) solution to the evolution equations for boost-invariant expansion with an equation of state $p\propto \varepsilon$ has $p\propto\varepsilon\propto \tau^{-4/3}$ and $s\propto 1/\tau$, meaning that if $\Phi\propto s/\tau$ then $\Phi$ grows faster for $\tau\rightarrow 0$ than $p$ does.  This means that at some early time, $\Phi$ must exceed $p$ making $P_\xi<0$.  Our results and the results of Ref.~\cite{Martinez:2009mf} confirm that the conclusions of this simple argument apply.  With the specific initial conditions that we have used --- namely with $\Phi=0$ initially --- negativity of $P_\xi$ can be avoided at any given $\eta/s$ by increasing $\tau_\Pi^\eta$ by a large enough factor.  This stretches out the initial transient, delaying the $\tau$ at which $\Phi\propto s/\tau$ is reached until late enough that $\Phi$ never exceeds $p$.  However, this resolution is specific to our (completely arbitrary) choice of the initial value of $\Phi$ --- we could have chosen $\Phi=4\eta/3\tau$ from the beginning --- and is therefore not actually a resolution.   The conclusion we should draw is simply that the hydrodynamic description, premised on local thermal equilibrium, must break down at early times and the negativity of $P_\xi$ is one sign that tells us before when we cannot go.  (Other evidence for the same qualitative and, essentially, quantitative conclusion has been developed in Refs.~\cite{Baier:2006um,Huovinen:2008te}.) For $\eta/s=1/4\pi$, initializing the hydrodynamic evolution equations at $\tau_0=0.5$~fm$/c$ as we are doing is appropriate.  For $\eta/s=2/4\pi$, choosing $\tau_0=0.5$~fm$/c$ is only appropriate for certain initial conditions (including $\Phi=0$), while choosing $\tau_0=1$~fm$/c$ is safe for more generic initial values of $\Phi$. For $\eta/s=3/4\pi$, we find that in order to avoid $P_\xi<0$ we must 
choose $\tau_0>(3-4)$~fm$/c$.    It would be interesting to pursue this line of reasoning further in the higher-dimensional hydrodynamic calculations of Refs.~\cite{Romatschke:2007mq,Song:2007fn,Dusling:2007gi,Song:2007ux,Luzum:2008cw,Song:2008si,Molnar:2008xj,Song:2008hj,Luzum:2009sb,Song:2009je,Heinz:2009xj}:  the same hydrodynamic calculations that yield information about the allowed values of $\eta/s$ via comparison to data from RHIC should at the same time constrain the earliest time at which a hydrodynamic description can be valid, via checking before what time these hydrodynamic evolutions feature negative pressure in some region of space.

The shear viscosity to entropy ratio $\eta/s$ becomes large at late times, in the hadron gas 
phase~\cite{Prakash:1993kd,Demir:2009qi}.  In our calculations with $\eta/s=1/4\pi$ and $\eta/s=2/4\pi$, the shear stress $\Phi$ is much less than $p$ at late times.   But, if we consider $\eta/s$ increasing at late times to $\eta/s\sim 1$, as appropriate at $T\sim 100-150$~MeV according to the calculations of Demir and Bass~\cite{Demir:2009qi}, we find $P_\xi$ coming close to going negative at the very late times corresponding to $T\sim 150$~MeV.  This would be worth further investigation, as a possible indicator of when the hydrodynamic description breaks down at late times, if not for the fact that freezeout in heavy ion collisions is expected to occur earlier than this. And, furthermore, we shall see that
including the effects of bulk viscosity can result in the breakdown of hydrodynamics also happening earlier, when $T\sim T_c$. To this we now turn.

\section{Effects of bulk viscosity}

We now wish to turn on a nonzero bulk viscosity $\zeta$ and study its effects on solutions of the evolution equations (\ref{EpsEquation}), (\ref{PhiEquation}) and, now, (\ref{PiEquation}).    We shall set $\eta$, $\tau_\Pi^\eta$ and $\lambda_1$ to their baseline values (\ref{KSS}), (\ref{tauPiBaseline}) and (\ref{lambda1Baseline}) throughout this Section.  In the case of $\tau_\Pi^\eta$ and $\lambda_1$ we do so because, as in~Section 2.5, we do not expect that changing their values would have significant consequences.  But, we have seen that varying $\eta/s$ is consequential. We are setting $\eta/s=1/4\pi$ in order to be conservative, in the following sense.  
 We shall focus on the question of whether the longitudinal pressure $P_\xi$ of (\ref{PLong}) goes negative.  Increasing $\eta$ increases $\Phi$, which makes a negative contribution to $P_\xi$. So, if we find that the bulk viscosity drives $P_\xi$ negative with $\eta/s$ set to $1/4\pi$, increasing $\eta/s$ would only make $P_\xi$ even more negative.

Both at low temperatures $T\ll T_c$ in the hadron gas~\cite{Chen:2007kx,FernandezFraile:2008vu,NoronhaHostler:2008ju,Wiranata:2009cz} and at very high temperatures $T\gg T_c$ where the quark-gluon plasma is weakly coupled~\cite{Arnold:2006fz} the bulk viscosity $\zeta$ is much smaller than the shear viscosity.
These calculations indicate that $\zeta/s$ rises as one approaches $T_c$ from both below and above.  And, 
if the crossover at $T_c$ were a second order phase transition, $\zeta$ would peak 
at $T_c$~\cite{Paech:2006st,Karsch:2007jc,Sasaki:2008fg}.  The general expectation that $\zeta/s$ may be significant near $T_c$ is supported by the analysis of Refs.~\cite{Kharzeev:2007wb,Karsch:2007jc} 
which uses sum rules to relate the bulk viscosity to (derivatives of) thermodynamic quantities calculated on the lattice (although this relation is subtle~\cite{Moore:2008ws,Romatschke:2009ng,CaronHuot:2009ns}) 
and by the analyses of Ref.~\cite{Meyer:2007dy} in which $\zeta$ itself is constrained via lattice calculations, albeit in QCD without quarks.   Both these approaches find $\zeta/s\sim 1 \gg \eta/s$ in a narrow range of temperatures very near $T_c$.  We shall ask whether such a peak in $\zeta/s$ can drive $P_\xi$ negative, triggering cavitation.

\subsection{Choosing $\zeta/s$ and $\tau_\Pi^\zeta$}

Determining $\zeta/s$ via lattice calculations of Euclidean correlation functions is challenging, and  the results obtained in Ref.~\cite{Meyer:2007dy} should not yet be seen as definitive~\cite{Moore:2008ws,Meyer:2008dq,Huebner:2008as}.  To give just one example of a difficulty~\cite{Moore:2008ws}, just as $\zeta$ peaks at a second order phase transition, so does the relaxation time $\tau_\Pi^\zeta$ --- due to the phenomenon of critical slowing down.  The Euclidean lattice calculations are sensitive to the ratio $\zeta/\tau_\Pi^\zeta$, making it hard to disentangle one from the other.  Note, however, that attributing a peak in $\zeta/\tau_\Pi^\zeta$ to $\zeta$ is conservative in the sense that by neglecting the rise in $\tau_\Pi^\zeta$ one underestimates the rise in $\zeta$.  
Unfortunately, other more technical challenges can go in the other direction~\cite{Meyer:2008dq,Huebner:2008as}.  Progress is nevertheless possible~\cite{Meyer:2008gt,Meyer:2009jp}.  The current state of affairs is that lattice calculations make a robust case for the existence of a peak in $\zeta/s$ at $T_c$, at least in QCD without quarks~\cite{Meyer:2009jp}, but they do not yet reliably determine the height of the peak.  We shall therefore parametrize the results of Ref.~\cite{Meyer:2007dy}, and vary the parameters over a considerable range.

In Ref.~\cite{Meyer:2007dy}, Meyer reports results for $\zeta/s$ at five values of the temperature, $T/T_c=$ 1.02, 1.24, 1.65, 2.22 and 3.22.  One way to parametrize his results is to write
\begin{equation}
\frac{\zeta}{s} = a \exp\left( \frac{T_c - T}{\Delta T} \right) + b \left(\frac{T_c}{T}\right)^2\quad{\rm for}\ T>T_c,
\label{ZetaFit}
\end{equation}
with $a$, $\Delta T$ and $b$ the parameters.  ($T_c$ is not a parameter in that QCD without quarks has a first order phase transition at a reliably determined $T_c$.  When we employ (\ref{ZetaFit}) together with the equation of state for QCD with quarks specified via (\ref{TraceAnomaly}), we shall set $T_c=190$~MeV.)  Meyer's results at the three higher temperatures, well above $T_c$, are consistent with $\zeta/s\propto 1/T^2$.  This is not surprising since the trace anomaly 
$(\varepsilon-3p)/(\varepsilon+p)$, which like $\zeta/s$ is a dimensionless measure of the breaking of conformal invariance, is $\propto 1/T^2$ at high temperature. (For example, see (\ref{TraceAnomaly}).)  If we set $a=0$ and fit $b$ to Meyer's central values of $\zeta/s$ at the three higher temperatures, we find 
\begin{equation}
b=0.061\ .
\label{bValue}
\end{equation}
Note that at these higher temperatures, Meyer gives the central values for $\zeta/s$ that we have used but his results remain consistent with $\zeta=0$. When we vary $b$, therefore, we should consider values as small as zero.
With $b$ chosen as in (\ref{bValue}) and with $a=0$, the curve (\ref{ZetaFit}) is far below Meyer's results at $T=1.02 T_c$ and $T=1.24 T_c$.  That is, there is no way to use simply $\zeta/s\propto 1/T^2$ to fit Meyer's high temperature results and his results at $T=1.02 T_c$ and $T=1.24 T_c$ --- where $\zeta/s$ has its peak.
The parameters $a$ and $\Delta T$ can then be chosen such that (\ref{ZetaFit}) passes directly through Meyer's central values of $\zeta/s$ at these two temperatures, yielding
\begin{equation}
a=0.901 \quad {\rm and} \quad \Delta T = \frac{T_c}{14.5}\ .
\label{aValue}
\end{equation}
The parameter $a$ controls the height of the peak in $\zeta/s$ and $\Delta T$ controls its width, and we shall vary both considerably.

Little is known about the value of $\tau_\Pi^\zeta$.  We shall use 
\begin{equation}
\tau_\Pi^\zeta=\tau_\Pi^\eta
\label{EqualTaus}
\end{equation} 
with $\tau_\Pi^\eta$ given in (\ref{tauPiBaseline}) as the baseline value for $\tau_\Pi^\zeta$. Although  there is no strong argument for this choice, it holds in one class of strongly coupled nonconformal fluids~\cite{Kanitscheider:2009as,Romatschke:2009kr}.\footnote{In kinetic theory, $\tau_\Pi^\zeta=\frac{5}{3}\tau_\Pi^\eta$~\cite{Huovinen:2008te}.}  On general grounds (i.e. as a manifestation of critical slowing down) and as in the specific strongly coupled nonconformal fluid studied in Ref.~\cite{Buchel:2009hv}, $\tau_\Pi^\zeta$ is expected to peak where $\zeta$ peaks, and we shall therefore check the effects of $\tau_\Pi^\zeta$ greater than in (\ref{EqualTaus}) by as much as a factor of 40.

\subsection{Boost invariant expansion with bulk viscosity}

Once one has picked $\zeta/s$ and $\tau_\Pi^\zeta$, the next step is to choose initial conditions.  We shall initialize at $\tau_0=0.5$~fm$/c$ and choose $\varepsilon$ as in Section 2.  We shall choose the initial shear stress $\Phi=0$ as in Section 2,  and we shall also choose the bulk stress $\Pi=0$.  When we then evolve the equations of motion (\ref{EpsEquation}), (\ref{PhiEquation}) and (\ref{PiEquation}), we find that $\Pi$ quickly evolves to its Navier-Stokes value (\ref{PiNavierStokes}) during an initial time that is controlled by $\tau_\Pi^\zeta$.    We shall always stop the evolution at the time when $T$ has dropped to $T_c$, since our parametrization of $\zeta/s$ in (\ref{ZetaFit}) is only valid for $T>T_c$. We expect $\zeta/s$ to drop rapidly below $T_c$, but even less is known about the shape of the $\zeta/s$ peak below $T_c$ than above it, so we simply stop the evolution when $T=T_c$ and ask whether by that time the longitudinal pressure $P_\xi$ has gone negative. 

In describing our results, we begin with $a=0$.  That is, we begin with no peak in $\zeta/s$, just with $\zeta/s\propto 1/T^2$.  If we choose $b$ as in (\ref{bValue}), suitable to describe the lattice results at $T>1.5 T_c$~\cite{Meyer:2007dy}, we find that the bulk viscosity has negligible effects.  Introducing the bulk viscosity changes the energy density by about 6\%. And, with $b$ as in (\ref{bValue}), we find that we must increase the shear viscosity $\eta/s$ to $1.8/4\pi$ in order to  see $P_\xi<0$ at early times --- whereas we saw in Section 2.6 that with no bulk viscosity this required $\eta/s=2.15/4\pi$.  
If we reduce $b$  relative to (\ref{bValue}), the effects of the bulk viscosity become even more negligible.
Even if we increase $b$ by a factor of 2 relative to (\ref{bValue}), we still find no qualitative consequences: the energy density changes by about 13\% and $P_\xi<0$ at early times requires $\eta/s>1.4/4\pi$.  Note that with $b$ greater than (\ref{bValue}) by a factor of two there is already a range of temperatures near $T_c$ where $\zeta>\eta$.  Increasing $b$ by another factor of two makes $\zeta>\eta$ over a wide range of temperatures, which is not supported by the lattice calculations.   Henceforth, we fix $b$ as in (\ref{bValue}), meaning that if $a$ were zero the bulk viscosity would not have interesting consequences.  

\begin{figure}[t]
\vskip-0.2in
\includegraphics[width=7.5cm,angle=270]{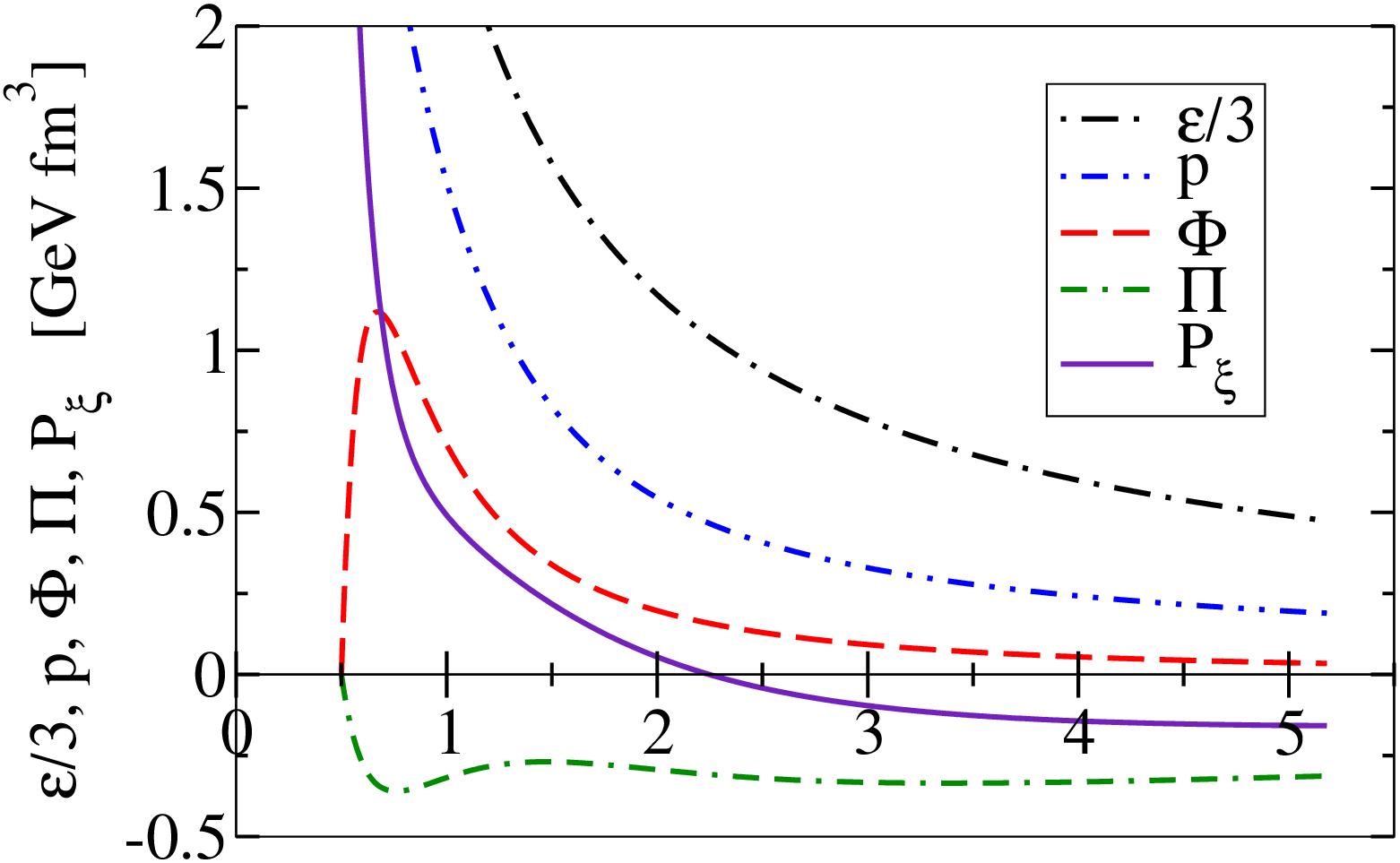}
\hskip-0.3in
\includegraphics[width=7.5cm,angle=270]{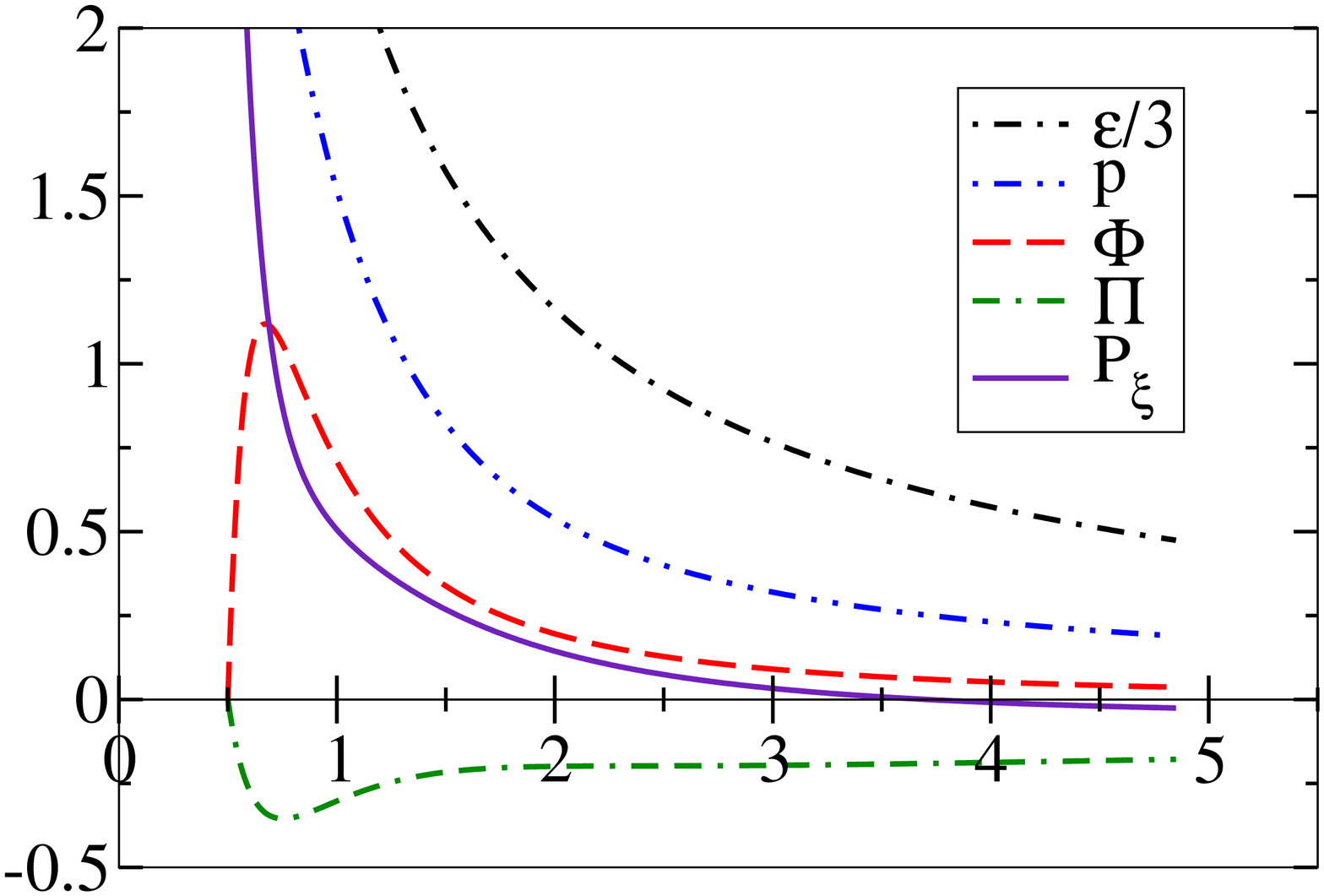}
\vskip-0.6in
\includegraphics[width=7.5cm,angle=270]{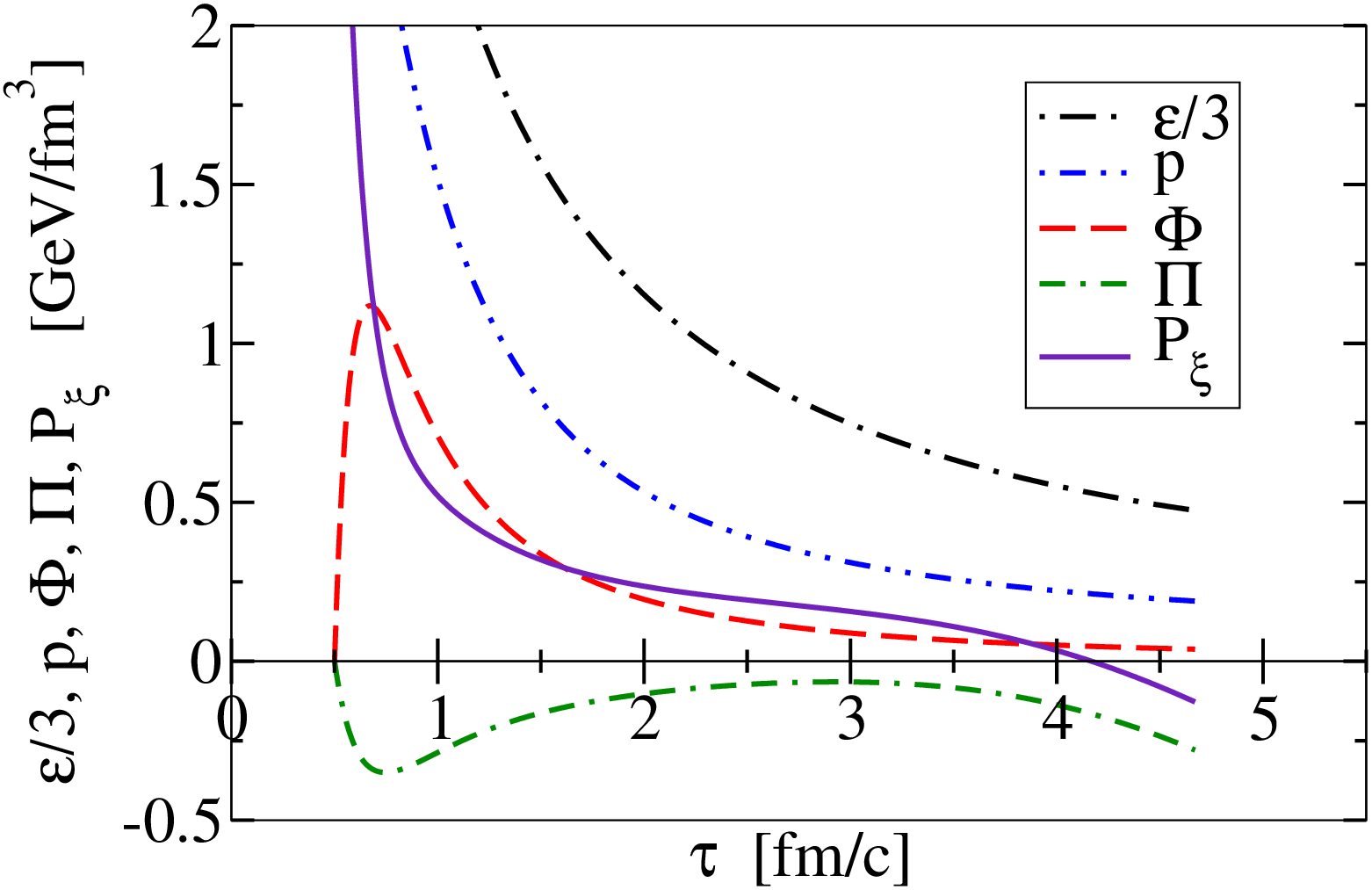}
\hskip-0.3in
\includegraphics[width=7.5cm,angle=270]{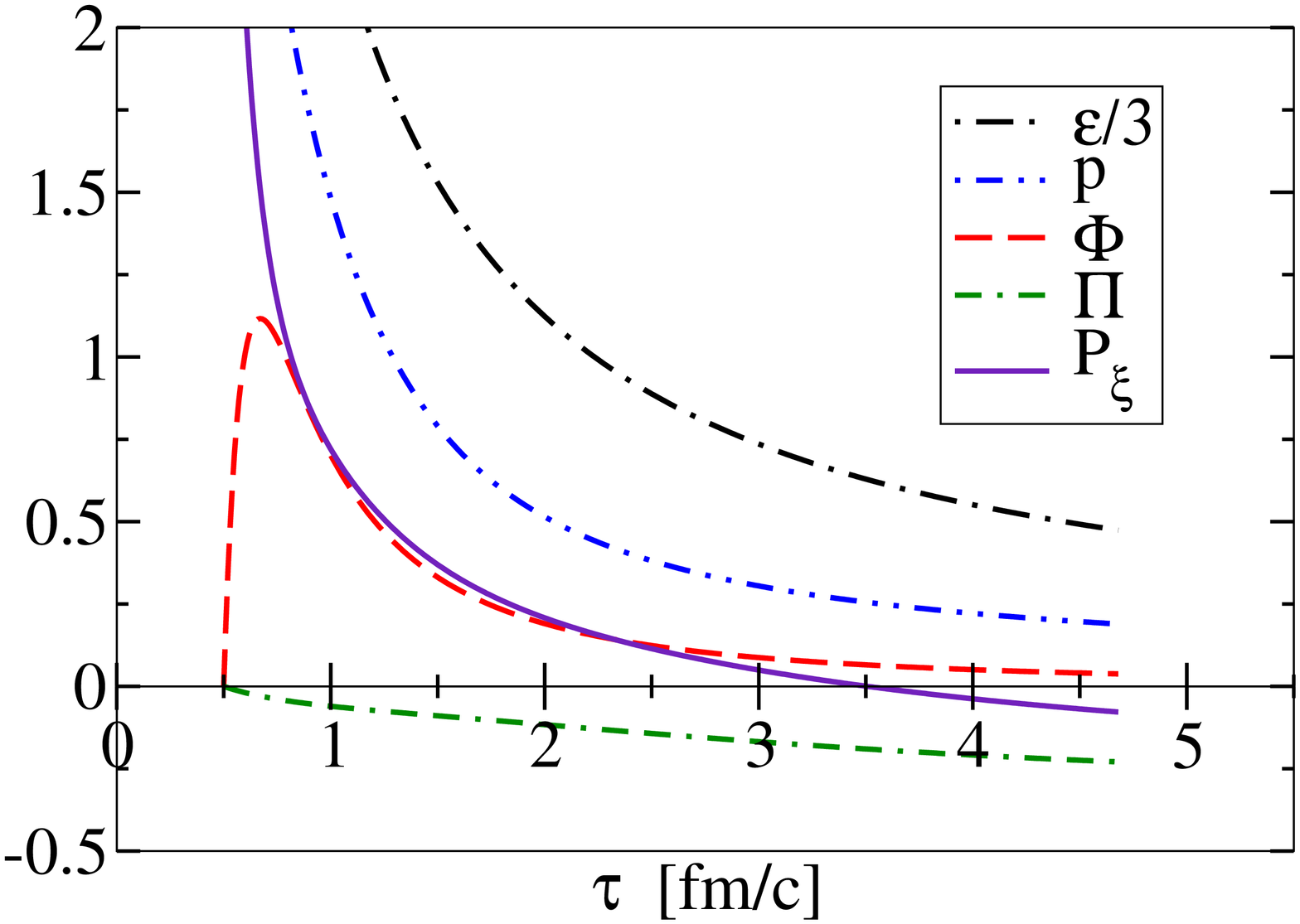}
\vskip-0.2in
\caption{
Consequences of a peak in the bulk viscosity near $T_c$.  We plot $\varepsilon/3$, $p$, the shear stress $\Phi$, the bulk stress $\Pi$, and the longitudinal pressure $P_\xi=p+\Pi-\Phi$ as functions of proper time $\tau$.  In all the panels, the curves end at the $\tau$ when $T=T_c=190$~MeV.
But, in all the panels, the peak in the bulk viscosity near $T_c$ has driven $P_\xi$ negative at some earlier time. 
After these times, the curves are not relevant because when $P_\xi$ reaches zero instead of continuing to expand smoothly the fluid would cavitate, falling apart into regions of fluid separated by regions of vacuum.
In the top-left panel, the bulk viscosity is as in ({\protect\ref{ZetaFit}}) with ({\protect\ref{bValue}}) and ({\protect\ref{aValue}}).  In the top-right panel, $a$ has been reduced relative to ({\protect\ref{aValue}}) by a factor of two --- the peak in the bulk viscosity near $T_c$ is half as high.  In the bottom-left panel, $a$ is as in ({\protect\ref{aValue}}) and $\Delta T$ has been reduced relative to ({\protect\ref{aValue}}) by a factor of four --- the peak in the bulk viscosity is four times as narrow.  In the bottom-right panel, $a$ and $\Delta T$ are as in ({\protect\ref{aValue}}) but $\tau_\Pi^\zeta=20\,\tau_\Pi^\eta$, whereas in all other panels $\tau_\Pi^\zeta=\tau_\Pi^\eta$.  In all the panels, the parameters $\eta$, $\tau_\Pi^\eta$ and $\lambda_1$ are set to their baseline values, as in Fig.~1a. 
(In the top-left panel, $P_\xi=0$ and cavitation occurs at $\tau=2.3$~fm$/c$ when $T=211$~MeV; in the top-right panel, at $\tau=3.7$~fm$/c$ when $T=195$~MeV; in the bottom-left panel, at $\tau=4.2$~fm$/c$ when $T=193$~MeV; in the bottom-right panel, at $\tau=3.5$~fm$/c$ when $T=197$~MeV.)  
}
\label{Fig:BulkPeak}
\end{figure}

We now investigate the consequences of the peak in $\zeta/s$ near $T_c$.  Let us begin by choosing $a$ and $\Delta T$ as in (\ref{aValue}), meaning that the peak in $\zeta/s$ has a height and width as indicated by Meyer's lattice results~\cite{Meyer:2007dy}.  
We illustrate the resulting evolution in the top-left panel of Fig.~\ref{Fig:BulkPeak}.  We see that with $a$ and $\Delta T$ as in (\ref{aValue}),  the rising bulk viscosity drives the longitudinal pressure $P_\xi$ negative when $T$ is  still well above $T_c$.   As we have described in Section 3.1, although there is good evidence for a peak in $\zeta/s$ near $T_c$, its height and width (which we are parametrizing by $a$ and $\Delta T$) are not well known.  So, we have explored for what values of these parameters $P_\xi$ is driven negative near $T_c$.  As the top-right panel of Fig.~\ref{Fig:BulkPeak} shows, upon reducing $a$ by a factor of 0.5 while keeping $\Delta T$ fixed we continue to find $P_\xi<0$.  In fact, we find that with $\Delta T$ as in (\ref{aValue}), the largest $a$ at which $P_\xi$ remains positive for all $T>T_c$, which we shall denote $a_{\rm stable}$,  is $a_{\rm stable}=0.37$, which is 41\% of the value (\ref{aValue}) indicated by the lattice calculations of Ref.~\cite{Meyer:2007dy}.\footnote{   
We have used the equation of state (\ref{TraceAnomaly}) and (\ref{LatticeEOS}) obtained from lattice QCD calculations throughout.
If instead we use the conformal equation of state $p=\varepsilon/3$ with no phase transition, then $a_{\rm stable}=0.95$. It is easy to see why a larger peak in the bulk viscosity is then required in order to drive the longitudinal pressure negative: in the vicinity of the peak in the bulk viscosity, $p=\varepsilon/3$ is significantly larger than $p$ in (\ref{LatticeEOS}) because (\ref{LatticeEOS}) describes a phase transition; because $p$ is larger, it takes more bulk stress to drive $P_\xi$ negative.  One measure of the robustness of our results is that even with the much larger  thermodynamic pressure $p=\varepsilon/3$, a peak in the bulk viscosity comparable to that indicated by the lattice calculations of Ref.~\cite{Meyer:2007dy} is sufficient to trigger cavitation.
}
How does changing the width $\Delta T$ modify this result?  If we increase $\Delta T$ by a factor of two relative to (\ref{aValue}), $a_{\rm stable}$ drops only slightly, to $0.33$.  If we increase it by another factor of two, making the peak in $\zeta/s$ four times wider than in Ref.~\cite{Meyer:2007dy}, $a_{\rm stable}$ drops to 0.22.
If we change $\Delta T$ in the other direction, decreasing it by a factor of two relative to (\ref{aValue}), meaning that we make the peak in $\zeta/s$ twice as narrow as in Ref.~\cite{Meyer:2007dy}, $a_{\rm stable}$ rises only very slightly, to $0.39$.    If we decrease $\Delta T$ by another factor of two, $a_{\rm stable}$ rises to $0.45$; if we decrease it by yet another factor of two --- making the peak in $\zeta/s$ eight times narrower than in Ref.~\cite{Meyer:2007dy} --- $a_{\rm stable}$ increases to 0.54.  In the bottom-left panel of Fig.~\ref{Fig:BulkPeak} we illustrate the case with $\Delta T$ reduced relative to (\ref{aValue}) by a factor of four, and $a$ unmodified, as in (\ref{aValue}).  Comparing this panel to the top-left panel, we see that reducing the width of the peak in the bulk viscosity by a factor of four delays the time at which $P_\xi$ goes negative, but does not significantly reduce the amount by which it goes negative.  This is consistent with the observation that $a_{\rm stable}$ is not much changed.  
We conclude that $a_{\rm stable}$ is 
quite insensitive to $\Delta T$ over a wide range of $\Delta T$,  ranging from 0.22 if $\Delta T$ is four times larger than (\ref{aValue}) to 0.45 if $\Delta T$ is four times smaller than (\ref{aValue}).   Over this wide range, $a_{\rm stable}$ is well below the value 0.90 indicated by the lattice calculations of Ref.~\cite{Meyer:2007dy}.

Note that all the results we have just quoted were obtained with $\eta/s=1/4\pi$. If we increase $\eta$, it will take a smaller $\zeta/s$ to push the longitudinal pressure negative.  For example, with $\Delta T$ as in (\ref{aValue}) if we increase $\eta/s$ to $2/4\pi$ this decreases the value of $a_{\rm stable}$ from 0.37 to 0.32.

It is also worth checking that the fact that $P_\xi$ is being driven negative really is due to the peak in the bulk viscosity, not to the $1/T^2$ component of $\zeta/s$ in (\ref{ZetaFit}).  To this end, we set $b=0$ (with $\eta/s=1/4\pi$ and $\Delta T$ as in (\ref{aValue})) and found that eliminating the $1/T^2$ component of $\zeta/s$ increased $a_{\rm stable}$, but only from 0.37 to 0.41.

The curves in all the panels in Fig.~\ref{Fig:BulkPeak} end at the $\tau$ when $T=T_c$ in the calculation, although they have ceased to be relevant earlier when $P_\xi$ reaches zero and cavitation occurs.  Note, however, that the $\tau$ at which $T=T_c$ is between 4.7~fm$/c$ and 5.2~fm$/c$ in all four panels, while in Fig.~\ref{Fig:baseline} $T$ reaches $T_c$ at 4.3~fm$/c$.  This is a small effect, but it can be made larger. As Kapusta discovered in Ref.~\cite{Kapusta:2008vb},  if $\zeta/s$ diverges at $T_c$, namely if $\zeta/s\propto 1/|T-T_c|^n$ for some positive power $n$, then the diverging bulk viscosity acts in the hydrodynamic equations so as to prevent $T$ from dropping below $T_c$: as $\tau\rightarrow\infty$ the equations show $T$ approaching $T_c$ from above but never reaching it.  The slight delay in the $\tau$ at which $T_c$ is reached in our calculations, as in Fig.~\ref{Fig:BulkPeak}, is the small residue of this effect when the peak in $\zeta/s$ is finite.  Note that the solution with diverging $\zeta/s$ and $T$ approaching $T_c$ from above is academic, since the $\tau\rightarrow\infty$ solution has $P_\xi$ negative. In fact, with a power-law divergent $\zeta/s$ the bulk shear is large enough to drive $P_\xi<0$ already at rather early times.  Cavitation occurs long before the asymptotic solution discovered in Ref.~\cite{Kapusta:2008vb} is reached.

Following Ref.~\cite{Fries:2008ts}, we have investigated the entropy produced according to the hydrodynamic equations that we have solved.  For example, comparing the calculation illustrated in the top-left panel of Fig.~\ref{Fig:BulkPeak}, in which $\zeta/s$ is as in (\ref{ZetaFit}) with (\ref{bValue}) and (\ref{aValue}), with the calculation illustrated in Fig.~\ref{Fig:baseline}, in which $\zeta=0$, we find that turning on a bulk viscosity with a peak as in Meyer's lattice results~\cite{Meyer:2007dy} increases the entropy by about 20\%. This suggests that entropy production due to the bulk viscosity is a modest effect, as was found previously in Ref.~\cite{Fries:2008ts} for cases in which the peak in $\zeta/s$ is not high enough to cause cavitation.   In our case, however, this result cannot be trusted because there is no way to use our boost-invariant calculation to estimate how much further entropy is produced upon cavitation.

We now consider the effects of increasing $\tau_\Pi^\zeta$, which is expected to be large where $\zeta$ peaks.  We will not attempt to model a time-varying $\tau_\Pi^\zeta$; instead, we ask what are the consequences of increasing $\tau_\Pi^\zeta$ relative to (\ref{EqualTaus}) at all temperatures.  If we increase $\tau_\Pi^\zeta$ by a factor of 10, keeping $\Delta T$ as in (\ref{aValue}), we find that $a_{\rm stable}$ decreases from 0.37 to 0.33.  If we increase $\tau_\Pi^\zeta$ by a further factor of two, meaning that it is 20 times greater than in (\ref{EqualTaus}), 
$a_{\rm stable}$ increases to 0.49.    If we increase $\tau_\Pi^\zeta$ by one further factor of two, making it 40 times greater than in (\ref{EqualTaus}), $a_{\rm stable}$ increases to 0.94, comparable to the $a$ in (\ref{aValue}).  In the bottom-right panel of Fig.~\ref{Fig:BulkPeak} we illustrate the case with $\tau_\Pi^\zeta=20\,\tau_\Pi^\eta$.  The bulk stress $\Pi$ changes more gradually as a function of time, and as a result $P_\xi$ does not go as far negative as in the top-left panel of Fig.~\ref{Fig:BulkPeak}, but the change is not dramatic and as a consequence $a_{\rm stable}$ is greater, but not by much.\footnote{After the first version of this paper appeared, Song and Heinz found that cavitation does not occur anywhere in their $(3+1)$-dimensional calculation if they use $\tau_{\Pi}^\zeta=(120~{\rm fm})(\zeta/s)$~\cite{Song:2009rh}, which is larger than (\ref{EqualTaus}) by a factor of several hundred in the vicinity of the peak in the bulk viscosity.  Our results are consistent with this.
}
This insensitivity to large changes in the value of $\tau_{\Pi}^\zeta$ is one indication that second order effects are still small at the time when cavitation occurs.  Another indication of this is the fact that $|\Pi|/(\varepsilon + p)$ is small, for example  less than 10\% at the time when cavitation occurs  in the top-left panel of Fig.~4.\footnote{As noted in Section 2.1,  once we break conformality by introducing a nonzero bulk viscosity  further second order terms can arise in both  (\ref{PhiEquation}) and (\ref{PiEquation})~\cite{Romatschke:2009kr}.   See Ref.~\cite{Denicol:2009zz} for one example.  We have confirmed that adding the second order terms considered in Ref.~\cite{Denicol:2009zz} has only negligible effects on our results.
}

We can summarize these results as follows:
\begin{itemize}
\item
With $a$ and $\Delta T$ as in (\ref{aValue}),  the peak in the bulk viscosity above $T_c$ drives the hydrodynamic evolution to negative $P_\xi$, indicating cavitation.
 \item
 Stable hydrodynamic evolution all the way down to $T=T_c$ requires that the peak in $\zeta/s$ near $T_c$ be less than a threshold that is one quarter to one half as high as the peak found in Ref.~\cite{Meyer:2007dy}; the threshold peak height is fairly insensitive to the width of the peak, for widths between one quarter and four times that found in Ref.~\cite{Meyer:2007dy}.
 \item
The effects of a peak in $\zeta/s$ near $T_c$ can be washed out by increasing $\tau_\Pi^\zeta$, the relaxation time for the bulk stress $\Pi$.  However, the increase must be by a very large factor. If $\tau_\Pi^\zeta$ is larger than $\tau_\Pi^\eta$ of (\ref{tauPiBaseline}) by a factor of 10 (or 20), results are little affected and the peak in $\zeta/s$ must still be reduced by about a factor of three (or two) relative to (\ref{aValue}) in order to obtain stable hydrodynamic evolution all the way down to $T=T_c$.  If 
$\tau_\Pi^\zeta$ is 40 times greater than $\tau_\Pi^\eta$, however, the effects of a bulk viscosity peak are washed out sufficiently that $P_\xi$  remains just barely positive even for a peak whose height and width are as in (\ref{aValue}).
\end{itemize}

We have focused on $P_\xi$ of (\ref{PLong}) rather than $P_\perp$ of (\ref{PPerp}) because the nonzero shear stress $\Phi$ implies that as the bulk stress $\Pi$ becomes increasingly negative, $P_\xi$ goes negative first, $P_\perp$ only later.  By the time $P_\perp$ goes negative in the calculation, the calculation has already broken down at the time when $P_\xi$ went negative, triggering cavitation. But, we see in all the panels in Fig.~\ref{Fig:BulkPeak} that $\Phi$ is quite small by the time $P_\xi$ goes negative, meaning that $P_\perp$ is already close to zero when $P_\xi$ reaches zero. It will be interesting to see, therefore, which component of the pressure goes negative first at which location in the fluid in a calculation that includes transverse expansion.

\section{Implications}

In thinking through the implications of our results, the first  possibility to consider is that the peak in 
$\zeta/s$ near $T_c$ in QCD with quarks is in fact not so high that cavitation results.  As we discussed in Section 3.1, the current lattice calculations of the height of the bulk viscosity peak in QCD without quarks have various caveats, meaning that the peak in this theory could be smaller (although it could just as well be larger) than is indicated by the results of Ref.~\cite{Meyer:2007dy}.   Furthermore, there are various indications that $\zeta/s$ is somewhat lower in QCD with quarks than in QCD without quarks.  At very high temperatures where the quark-gluon plasma is weakly coupled, if one compares the two theories at a fixed small value of the QCD coupling, say $\alpha_{\rm QCD}=0.2$, one finds that $\zeta/s$ in QCD with three flavors of quarks is about 56\% of that in quarkless 
QCD~\cite{Arnold:2006fz}.   At these high temperatures, $\zeta/s$ is smaller than $10^{-3}$ in value, so this comparison can give only rough guidance to how the height of the peak near $T_c$ will change when quarks are introduced, but it does suggest that it will decrease.  A second argument is simply that the transition in QCD with quarks is a crossover whereas that in quarkless QCD is first order, and if adding quarks smooths out the transition then it is reasonable to guess that it will also round off the peak in $\zeta/s$.  The magnitude of this effect can be guessed by looking 
at $(\varepsilon-3 p)/(\varepsilon + p)$ which, like $\zeta/s$, is a dimensionless measure of the breaking of conformality.  In quarkless QCD, $(\varepsilon-3 p)/(\varepsilon+p)$ peaks at a value of 0.85~\cite{Boyd:1996bx} while in the equation of state (\ref{TraceAnomaly}) for QCD with quarks, $(\varepsilon-3 p)/(\varepsilon+p)$ peaks at  0.53.   So, both this argument and the comparison to what happens at very high temperatures suggest that the peak in $\zeta/s$ in QCD with quarks is (very roughly) about half as high as in (\ref{ZetaFit}) with $a$ as in (\ref{aValue}).  Taking our results at face value, this would put it just above $a_{\rm stable}$, meaning that the peak in $\zeta/s$ would trigger cavitation very close to $T_c$.  The uncertainties are large and it could certainly be that the peak height ends up lower than $a_{\rm stable}$, and no cavitation occurs near $T_c$.  The previous studies of the effects of the peak in the bulk viscosity in Refs.~\cite{Fries:2008ts,Song:2008hj,Song:2009je} have explored the consequences of peaks that are not high enough to cause cavitation.\footnote{It is worth noting that in the examples of nonconformal plasmas in which the authors of Refs.~\cite{Gubser:2008yx,Kanitscheider:2009as,Buchel:2009bh,Gursoy:2009kk} have been able to compute $\zeta/s$ via gauge/gravity duality, a peak in $\zeta/s$ is seen but it is not large enough to cause cavitation.  For example, in the model of Ref.~\cite{Kanitscheider:2009as} the ratio $\zeta/\eta$ is given by $2(\frac{1}{3}-c_s^2)$, with $c_s$ the speed of sound, and is therefore everywhere less than $2/3$.
In contrast, in the example analyzed in Ref.~\cite{Buchel:2008uu} via gauge/gravity duality $\zeta$ can be $\gg \eta$ and $\zeta/s$ rises comparably high to the peak found in Meyer's lattice calculations~\cite{Meyer:2007dy}, more than high enough to trigger cavitation.  At present, these calculations taken together therefore support the existence of a peak in $\zeta/s$ but do not provide sufficient guidance regarding its height.}
If this is the path that nature chooses, then hydrodynamic calculations can be followed down to temperatures below $T_c$, and it becomes interesting to investigate the possibility of cavitation at a lower temperature, driven by the rising shear viscosity of the hadronic phase at low temperatures~\cite{Prakash:1993kd,Demir:2009qi}. 

It is more interesting to consider the possibility that the peak in $\zeta/s$ near $T_c$ in QCD with quarks {\it is} large enough to cause the expanding fluid produced in heavy ion collisions to cavitate when it cools through $T\sim T_c$, as our $1+1$-dimensional calculations indicate.  There are a variety of aspects of the observed phenomenology of heavy ion collisions that give some support to this possibility:   
\begin{itemize}
\item
The possibility that the peak in the bulk viscosity near $T_c$ can cause the fluid to fragment, and then freezeout, has been considered previously in Refs.~\cite{Torrieri:2007fb}.  These authors suggest that data on two particle momentum correlations (the HBT effect) in heavy ion collisions at RHIC can be understood if the hadrons in the final state come from such fragments. They have also suggested other experimental observables in Ref.~\cite{Melo:2009xh}.
\item
One of the workhorses of heavy ion phenomenology is the statistical hadronization model, reviewed in Ref.~\cite{Becattini:2009sc}, which has been used successfully to describe the ratios among the yields of many different hadrons using a few parameters including the chemical freezeout temperature and baryon chemical potential.  One of the conceptual underpinnings of this model, described in Ref.~\cite{Becattini:2009sc} and going back to the original formulation of Hagedorn~\cite{Hagedorn:1965st},  is the assumption that high energy collisions give rise to multiple clusters --- colorless, extended, massive objects --- which then hadronize statistically (meaning that all hadronic final states consistent with conservation laws are equally likely).  Most work in this context has focused on the statistical hadronization; the dynamics that results in the generation of the clusters in the first place has received less attention.  This dynamics is no doubt complex, and may be quite different in hadron-hadron and heavy ion collisions.  Our work suggests that in the case of ultrarelativistic heavy ion collisions, in which a hydrodynamic description in terms of an expanding fluid with $T>T_c$ is appropriate in the early stages of the collision, the clusters required by the statistical hadronization model may arise via cavitation, and this cavitation may be triggered by the peak in $\zeta/s$ in the vicinity of $T_c$.  Perhaps this could be an explanation of why the chemical freezeout temperatures extracted via the use of the statistical hadronization model seem to be in the vicinity of $T_c$~\cite{RHIC}.
\item
In Refs.~\cite{Broniowski:2005ae}, Broniowski et al have explained the event-by-event fluctuations in the mean transverse momentum of the particles produced in heavy ion collisions at RHIC via the same assumption that underlies the statistical hadronization model, namely that hadronization is preceded by the material produced in a heavy ion collision falling apart into clusters, each of which then yield  6 to 15 hadrons when they hadronize.
\item
In Ref.~\cite{Alver:2008gk}, the PHOBOS collaboration provides evidence (from two-particle correlations in pseudorapidity and azimuthal angle) that the hadrons in the final state produced in heavy ion collisions at RHIC come from clusters that decay into  3 to 6 charged hadrons, meaning 5 to 9 hadrons in all.
\end{itemize}
As we discussed in Section 1, it is pleasing to have a means by which a hydrodynamic calculation can predict its own break down.   The peak in the bulk viscosity near $T_c$ can provide a simple and elegant means: if this peak is high enough --- as we have quantified --- it drives the longitudinal pressure to zero at which point the fluid cavitates.  The phenomenological evidence in support of the notion that hadronization in ultrarelativistic heavy ion collisions is preceded by cavitation, with the fluid fragmenting into droplets that play the role of the clusters which have long been employed in the statistical hadronization framework, is perhaps not yet overwhelming.  But, together with our investigation, it certainly seems sufficient to take this possibility seriously.

There are many avenues open for further investigation:
\begin{itemize}
\item
An investigation of the effects of other possible second order terms that can arise in the hydrodynamic equations for a nonconformal fluid, as well as third order terms, would be desirable.    Little is known about the coefficients of such terms. But, although they will have quantitative effects, given the insensitivity of our results to large variations in $\tau_\Pi^\eta$, $\lambda_1$ and $\tau_\Pi^\zeta$ there is no reason to expect qualitative effects.
\item
A more interesting direction to pursue is to repeat our study using a $3+1$-dimensional hydrodynamic code, describing both longitudinal and transverse expansion.   This will make the determination of the height of the peak in $\zeta/s$ that is needed in order to trigger cavitation when the fluid cools through $T_c$ more quantitative.  And, having a fluid whose energy density varies with tranverse position will raise new questions and open new possibilities. For example, 
cavitation should occur much earlier at the (cooler) edges of the collision 
region, since $T\sim T_c$ 
there earlier~\cite{Song:2009rh,HeinzPrivate}.  Cavitation should start at the edges and move inward, just as hadronization has long been understood to do.
\item
It is important to investigate whether if freezeout and hadronization are triggered by cavitation near $T_c$ this modifies the extraction of $\eta/s$ via comparison between data and hydrodynamic calculations of the anisotropic expansion of the fluid produced in collisions with a nonzero impact parameter.  The effect of the physics of cavitation near $T_c$ on this comparison may prove minimal, since the anisotropic flow is generated early in the collision, when the hot fluid is still azimuthally anisotropic in shape. This means that the anisotropic flow is generated well before cavitation is triggered, when the bulk viscosity is still small compared to the shear viscosity.
\item
From the point of view of understanding the observable, and perhaps observed, phenomenology of cavitation, the most important question is the determination of the size distribution of the droplets formed when the fluid cavitates.  Work in this direction can be found in Refs.~\cite{Torrieri:2007fb}.    A from first principles determination of the size distribution will be challenging: for one, it will require determining the surface tension associated with the interface between expanding quark-gluon plasma with $T\sim T_c$ and vacuum.  If this surface tension is small, small droplets will be favored.   The requirement that the droplets must be color singlets will require rearrangement of color within 
$\sim \Lambda_{\rm QCD}^{-1}$ 
of the surface as droplets separate during cavitation.  Although hard to quantify, this can be thought of as a contribution to the surface tension which sets a limit on the smallness of the droplets that is of order  a few times $\Lambda_{\rm QCD}^{-1}$.   It is then interesting to note that a spherical droplet with a radius of 1 fm that has the energy density obtained from the equation of state (\ref{TraceAnomaly}) at $T=T_c$ contains about 6 GeV of energy, which is in the right ballpark to explain the PHOBOS data~\cite{Alver:2008gk} mentioned above.  This suggests that the surface tension is small enough that cavitation yields many small droplets.  The picture to have in mind is that as the quark-gluon plasma produced in an ultrarelativistic heavy ion collision expands and cools through $T\sim T_c$, the fluid falls apart into a mist of hundreds of small droplets, each of which later hadronizes as in the statistical hadronization model.   Cavitation into a mist of small droplets which then become a hadron gas is not likely to have dramatic observable consequences 
\end{itemize}

\medskip
\section*{Acknowledgments}
\medskip

We acknowledge very helpful conversations with Ulrich Heinz, Harvey Meyer,  Tomoi Koide, Gunther Roland, Paul Romatschke, Huichao Song, Misha Stephanov and Derek Teaney.
NT is grateful to the Research Science Institute of the Center for Excellence in Education for supporting his research.
This research was
supported in part by the DOE Office of Nuclear 
Physics under contract \#DE-FG02-94ER40818.


\end{document}